# The effect of undulations on Spontaneous Braid formation

Dominic Lee[1,a.)]

[1]Department of Chemistry, Imperial College London, SW7 2AZ, London, UK

## Abstract

This paper extends on a recent work where it was shown that forces dependent on the helical structure may cause two DNA molecules to spontaneously braid [R. Cortini et Al, Biophys. J. **101,** 875 (2011)]. Here, bending fluctuations of DNA centre lines about the braid axis are incorporated into the braiding theory. The free energy of the pair of molecules is recalculated and compared to its value without incorporating undulations. We find that the loss of configurational entropy due to confinement of the molecules in the braid is rather high. This contribution to the Free Energy pushes up the amount of attraction needed for spontaneous braiding due to helix dependant forces. The theory will be further developed for plectonemes and braids under mechanical forces, in later work.

## 1. Introduction

In a recent paper [1], the possibility of two DNA molecules spontaneously braiding through helix specific interactions, was investigated. It was found that, as DNA is a right handed, helix specific interactions favour a left handed braid. This was also argued from simulation data and x-ray scattering in the work of Timsit and Varnai [2]. One of the limitations of Ref. [1] was that it was a ground state calculation in terms of the bending degrees of freedom, so could not estimate the confinement entropy. This current work attempts to address this issue by incorporating such undulations, thereby developing a more complete theory for molecular braids with helix specific interactions.

In the past, two statistical mechanical theories have been developed to deal with undulations in braids and plectonemes [3, 4]. Both of these theories rely on interaction theories that treat the molecules as uniformly charged rods. On the other hand, there have been developed interaction theories [5,6,7,8] that incorporate helical charge distributions to describe forces between parallel molecules. The most important qualitative feature of these theories is that they depend on how the molecules are azimuthally orientated about their long axes. The Kornyshev-Leikin (KL) theory [5, 6] deals with a mean-field electrostatic theory that assumes a bulk dielectric response. Conversely, in the work of Ref [7], the effect of helical charge distributions was investigated in the limit of strong correlations between ions in solution about helically charged molecules. Last of all, Ref [8] considered corrections to the KL theory due to ion correlations and steric effects. In these

---

[a.)] Electronic Mail: domolee@hotmail.com

works a significant azimuthal dependence of the interaction potential was found, at close enough distances, and at certain values of the various model parameters. These theories also give rise to possibility of spontaneous braid formation from the appearance of a chiral torque due to the helical nature of the molecules [1,2].

For a complete description of braids with helix dependant forces - a fully consistent theory- the statistical mechanics describing undulation effects [3,4] needs to be modified. This needs to take account of non-trivial effects for the molecular twisting degrees of freedom as well as a chiral (braiding) torque [1,2], arising from such forces. In assemblies of DNA, a full statistical mechanical theory of undulations and twisting was developed [9] that took account of helix dependant forces and steric confinement. This theory built on the works of [10,11,12,13] that dealt with undulations and confinement of the molecules. With some modification these developments could be applied to braids.

As a first step to achieving this goal, we consider braids formed by only helix dependent forces, unconstrained by topology or mechanical forces. In developing such a theory, we want to see how undulations affect the results of Ref. [1]. The helix dependant interaction theory we choose is the KL theory, but the whole approach can be modified to any interaction theory where the helical shape of the molecules is important. It could be easily adapted to the strong correlation theory of Ref. [7] or an empirical theory constructed from simulation results, such as those of Ref. [2]. In later work, we hope to extend this statistical mechanical treatment to braids under the additional influences of topology and mechanical forces.

Could this really matter in the determination of the equilibrium properties of a molecular braid or plectoneme under certain conditions? For DNA assemblies, in the presence of condensing agents, and toriodal structures formed by DNA there is evidence to suggest that helical structure does indeed matter. The decay lengths of the forces between molecules from experiments [14,15,16] agree well with the KL theory [5,6,9], where these lengths arise from the helical structure, and their magnitude is fitted reasonably well by the results of Ref. [9]. On top of that, there is evidence of azimuthal order [17,18,19], a preferred orientation for each DNA molecule or segment about its long axis [9]. Therefore, it does not seem unreasonable to expect, in certain cases, that helical structure might matter in the formation of DNA braids and plectonemes and those formed by other charged, helical molecules. However, this still remains to be seen experimentally.

The main paper is divided into three further sections. In the theory section we discuss how the braid geometry can be described mathematically, using an approach similar to [20]. Next, using results of the supplemental material and electrostatic calculations [1,21], we write down a partition function that describes the thermal fluctuations of the braid and how steric effects can be estimated using a similar approach to [9]. Last of all, we outline a variational approximation using the Gibbs-Bogoliubov inequality and present an expression for the free energy in terms of the variational parameters, each of which has clear physical meaning. In the results section we calculate the free energy, parameters to with average braid structure, and quantities that characterize the size of fluctuations about the mean braid structure. We do this calculation for a braid formed of two homologous sequences and two completely random sequences, which we compare against the case where braid undulations are not included. Finally in the discussion section we discuss the significance the results, the limitations of the theory and point to new work.

## 2. Theory

### 2.1 Specifying fluctuating braid geometry

We will consider a fluctuating braid where the braid axis is still assumed to be straight, undulations of that axis will be considered in a later work. This is on a par with the theories of [3,4]. The braid axis can be written as $\mathbf{r}_A(z) = z(s)\hat{\mathbf{k}}$, the molecular centre lines of the molecules (labelled $1$ and $2$) can be written as

$$\mathbf{r}_1(s) = z(s)\hat{\mathbf{k}} - \frac{R(s)\hat{\mathbf{d}}(s)}{2}, \quad \mathbf{r}_2(s) = z(s)\hat{\mathbf{k}} + \frac{R(s)\hat{\mathbf{d}}(s)}{2}. \tag{2.1}$$

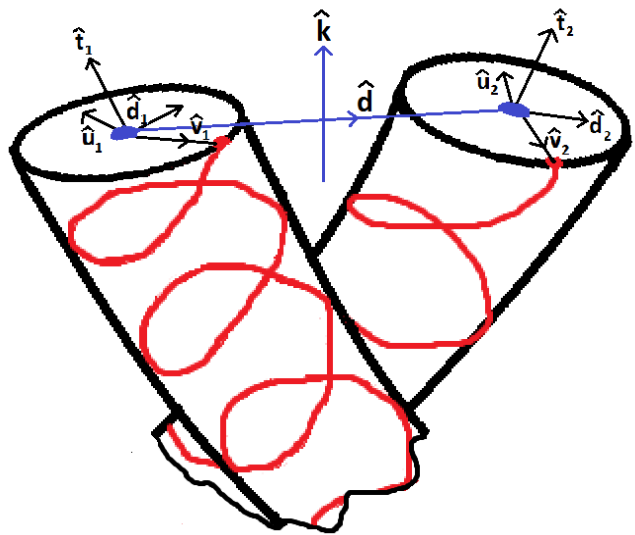

Fig 1 Diagram showing part of the braid formed of two molecules. Here the red line traces out the position of the minor groove in a distorted helical pattern due to thermal fluctuations and base pair clashes. The blue dots represent the DNA centre lines. A blue line of length $R(s)$ connects the two centre lines that is perpendicular to the braid axis, pointing along the z-axis. The unit vector $\hat{\mathbf{d}}(s)$ points along this line connecting the molecules. Two braid frames can describe the orientation of the DNA cross-sections (at fixed values of $s$)

relative to the line connecting the two centre lines, described by the basis sets $\{\hat{\mathbf{d}}_1,\hat{\mathbf{n}}_1,\hat{\mathbf{t}}_1\}$ and $\{\hat{\mathbf{d}}_2,\hat{\mathbf{n}}_2,\hat{\mathbf{t}}_2\}$. The orientation of the minor grooves can be described with respect to these frames by Eq. (2.5).

$R(s)\hat{\mathbf{d}}(s)$ is a vector that connects the two molecular centrelines and is only perpendicular to the tangent vectors $\hat{\mathbf{t}}_1(s) = \mathbf{r}'_1(s)$ (the prime here refers to differentiation with respect to argument), $\hat{\mathbf{t}}_2(s) = \mathbf{r}'_2(s)$ when $R'(s) = 0$ (see Fig 1); but, it is always perpendicular to the tangent vector of the braid axis $\hat{\mathbf{k}}$. Here, $s$ is a unit arc-length coordinate that runs from $-L/2$ to $L/2$, where $L$ is the contour length of the molecule. By constructing other unit vectors

$$\hat{\mathbf{n}}_1(s) = \frac{\hat{\mathbf{t}}_1(s) \times \hat{\mathbf{d}}(s)}{\left|\hat{\mathbf{t}}_1(s) \times \hat{\mathbf{d}}(s)\right|}, \qquad \hat{\mathbf{d}}_1(s) = \hat{\mathbf{n}}_1(s) \times \hat{\mathbf{t}}_1(s), \qquad (2.2)$$

$$\hat{\mathbf{n}}_2(s) = \frac{\hat{\mathbf{t}}_2(s) \times \hat{\mathbf{d}}(s)}{\left|\hat{\mathbf{t}}_2(s) \times \hat{\mathbf{d}}(s)\right|}, \qquad \hat{\mathbf{d}}_2(s) = \hat{\mathbf{n}}_2(s) \times \hat{\mathbf{t}}_2(s), \qquad (2.3)$$

we construct two local orthogonal frames called the *braid frames* [20] spanned by the basis sets $\{\hat{\mathbf{d}}_1,\hat{\mathbf{n}}_1,\hat{\mathbf{t}}_1\}$ and $\{\hat{\mathbf{d}}_2,\hat{\mathbf{n}}_2,\hat{\mathbf{t}}_2\}$. One should note only when $R'(s) = 0$ is $\hat{\mathbf{d}}_1(s) = \hat{\mathbf{d}}_2(s) = \hat{\mathbf{d}}(s)$, otherwise all three vectors point in different directions. We define the tilt angle $\eta(s)$ through the following relation between tangent vectors

$$\hat{\mathbf{t}}_1(s) \cdot \hat{\mathbf{t}}_2(s) = \cos\eta(s). \qquad (2.4)$$

In the braid frames we can describe the orientation of helix (for DNA, the position of the minor groove) through the vectors (see Fig 1)

$$\hat{\mathbf{v}}_1(s) = \cos\phi_1(s)\hat{\mathbf{d}}_1(s) + \sin\phi_1(s)\hat{\mathbf{n}}_1(s), \qquad \hat{\mathbf{v}}_2(s) = \cos\phi_2(s)\hat{\mathbf{d}}_2(s) + \sin\phi_2(s)\hat{\mathbf{n}}_2(s). \qquad (2.5)$$

**2.2 Constructing the full energy functional**

We, therefore, have a set of functions $\{R(s),\eta(s),\bar{\phi}(s),\Delta\Phi(s)\}$ that describe the twisting and bending fluctuations of the two molecules forming the braid, where $\bar{\phi}(s) = \phi_1(s) + \phi_2(s)$ and $\Delta\Phi(s) = \phi_1(s) - \phi_2(s)$. In Appendix A of the Supplemental Material we derive a total energy functional between two molecules forming a braid that is the sum of an elastic energy, helix specific interaction energy and a steric term, namely

$$E_T[R(s),\eta(s),\Delta\Phi(s)] = E_{elast}[R(s),\eta(s),\Delta\Phi(s)] + E_{int}[R(s),\eta(s),\Delta\Phi(s)] + E_{st}[R(s)]. \qquad (2.6)$$

For relatively small tilt angles, $\bar{\phi}(s)$ is unimportant and can be effectively decoupled from the problem provided that the helical persistence length (a measure of the rigidity of the helix against distortions, which is a combination or torsional and stretching rigidities [9]) $l_p^h$ is sufficiently large (see Appendix A of Supplemental Material). The helical persistence length is given by $l_p^h = C_t C_s / (C_s + g^2 C_t) k_B T$, where $C_t$ and $C_s$ are the twisting and stretching rigidities, respectively. We estimate $l_p^h \approx 400 \text{Å}$, for DNA, based on a torsional rigidity $C_t / k_B T \approx 1000 \text{Å}$ measured in recent twisting experiments [22] and a value of $C_s \approx 10^{-4}$ dyn [6] .

The elastic energy can approximated as (Appendix A of Supplemental Material), provided that $R'(s) \ll \cos \eta(s)$,

$$E_{elast}[R(s), \eta(s), \Delta\Phi(s)] \approx k_B T \int_0^{L_B} ds \mathcal{E}_{elast}\left(\Delta\Phi'(s), R''(s), R'(s), R(s), \eta'(s), \eta(s)\right), \tag{2.7}$$

where

$$\mathcal{E}_{elast}\left(\Delta\Phi'(s), R''(s), R'(s), R(s), \eta'(s), \eta(s)\right) = \left[\frac{l_p^h}{4}\left(\frac{d\Delta\Phi(s)}{ds} - \sigma_H \Delta\Omega(s)\right)^2 + \frac{l_p^b}{4}\left(\frac{d^2 R(s)}{ds^2}\right)^2 \right.$$
$$\left. + \frac{l_p^b}{4}\left(\frac{d\eta(s)}{ds}\right)^2 + \frac{l_p^b(1-\cos\eta(s))^2}{R(s)^2} - \frac{l_p^b(1-\cos\eta(s))}{2R(s)^2}\left(\frac{dR(s)}{ds}\right)^2 + \frac{3l_p^b \sin\eta(s)}{2R(s)}\left(\frac{dR(s)}{ds}\right)\left(\frac{d\eta(s)}{ds}\right)\right].$$

(2.8)

The parameter $l_p^b$ is the bending persistence length (for DNA, we take $l_p^b \approx 500 \text{Å}$). Here, $\sigma_H$ is a factor that depends on whether DNA molecules are homologous or non-homologous to each other in the braid; $\sigma_H = 0$, for homologous molecules and $\sigma_H = 1$ for non-homologous molecules. For other helical molecules that form regular helices in their ground states one can simply set $\sigma_H = 0$. When $\sigma_H = 1$ the function $\Delta\Omega(s)$ becomes important, a random Gaussian field that represents the mismatch between two non-homologous helices of two non-homologous DNA molecules due to the difference between them in base pair sequence. The field $\Delta\Omega(s)$ is uncorrelated so that

$$\langle\Delta\Omega(s)\Delta\Omega(s')\rangle_{\Delta\Omega} = \frac{2\langle h\rangle^2}{\lambda_c^{(0)}}\delta(s-s'), \tag{2.9}$$

where $\langle h \rangle$ is the average rise between base pairs and $\lambda_c^{(0)}$ is the intrinsic contribution to the helical coherence length, which describes the rate at which two non-interacting helices fall out of register with each other (for more of a discussion of the physics see [23]). Here, the averaging bracket $\langle\ldots\rangle_{\Delta\Omega}$ corresponds to an ensemble average over all the realizations of $\Delta\Omega(s)$.

If $\sin\eta(s)$ remains small and $R'(s) \ll 1$, we can use results for the electrostatic energy of a braid with a straight axis by simply replacing constant $R$ with $R(s)$, as well as making $\Delta\Phi$ and $\eta$ both $s$ dependent (for a justification of this procedure for $\Delta\Phi$ and $\eta$, and how to extend the electrostatic theory of the braid, see Ref. [21]).

$$E_{int}[R(s),\eta(s),\Delta\Phi(s)] = k_B T \int_0^{L_B} ds \mathcal{E}_{int}\left(\Delta\Phi(s),R(s),\eta(s)\right), \qquad (2.10)$$

$$\mathcal{E}_{int}\left(\Delta\Phi(s),R(s),\eta(s)\right) = \left(E_{img}(R(s)) + \sum_{n=0}^{2}\left\{E_0^{(n)}(R(s)) + \sin\eta(s)E_1^{(n)}(R(s))\right\}\cos n\Delta\Phi(s)\right). \qquad (2.11)$$

For the KL theory for the braid [1] the interaction coefficients are

$$E_{img}(R) = -\frac{2l_B}{l_c^2}\sum_{n=-\infty}^{\infty}\sum_{j=-\infty}^{\infty}\frac{K_{n-j}(\kappa_n R)K_{n-j}(\kappa_n R)}{(\kappa_n a K_n'(\kappa_n a))^2}\frac{I_j'(\kappa_n a)}{K_j'(\kappa_n a)}\zeta_n^2, \qquad (2.12)$$

$$E_0^{(0)}(R) = \frac{2l_B(1-\theta)^2}{l_c^2}\frac{K_0(\kappa_D R)}{(\kappa_D a K_1(\kappa_D a))^2}, \qquad (2.13)$$

$$E_0^{(n)}(R) = \frac{4l_B}{l_c^2}\frac{(-1)^n K_0(\kappa_n R)}{(\kappa_n a K_n'(\kappa_n a))^2}\zeta_n^2 \quad \text{for } n\neq 0, \qquad (2.14)$$

$$E_1^{(n)}(R) = \frac{4l_B n^2 a\bar{g}}{l_c^2}\frac{(-1)^n K_1(\kappa_n R)}{(K_n'(\kappa_n a))^2(\kappa_n a)^3}\zeta_n^2, \qquad (2.15)$$

where $\kappa_n = \sqrt{\kappa_D^2 + n^2\bar{g}^2}$, $\bar{g} = 2\pi/H$, and the form factors $\zeta_n$ are the helical Fourier components (modes) of the molecular surface charge density [6]. For DNA, a simple model of counter-ion binding and condensation gives the form factor

$$\zeta_n(f_1,f_2,\theta) = \delta_{n,0}\theta(1-f_1-f_2) + \theta\left[f_1 + (-1)^n f_2\right]-\cos n\tilde{\phi}_s, \qquad (2.16)$$

for other possible form factors see [6]. Here, $l_c$ is the mean separation per unit charge (for DNA $l_c = \langle h\rangle/2 \approx 1.7\text{Å}$), $l_B$ is the Bjerrum length (taken to be $7\text{Å}$ at room temperature). The parameters $\kappa_D$, $a$ and $H$ are the inverse Debye screening length, the effective cylinder radius and average helical pitch. For DNA, the values of $a \approx 11.5\text{Å}$ and $H \approx 33.8\text{Å}$ are taken, along with $\tilde{\phi}_s \approx 0.4\pi$ that is the half width of the minor groove. The functions $I_n(x)$ and $K_n(x)$ are modified functions of the first and second kinds of order $n$ and $I_n'(x)$ and $K_n'(x)$ are their respective derivatives with respect to argument. The parameter $\theta$ is the fraction of the molecule neutralized by counter-ions and, for DNA, $f_1$ and $f_2$ are the relative proportions of ions localized in the minor and major grooves, respectively.

We should emphasize that the overall form of Eq. (2.11) is not restricted to the results, Eqs. (2.12)-(2.16), given by the KL theory. Its form can be argued heuristically from geometric and symmetry considerations for helix specific interactions, provided that $R(s)$, $\sin\eta(s)$ and $\Delta\Phi(s)$ vary slowly compared the inverse decay ranges of the interaction. Indeed, the coefficients $E_0^{(n)}(R)$ and $E_1^{(n)}(R)$ may be fitted to simulations or determined from an alternative theory where helical symmetry is important.

The simplest way to model the steric term is to assume that molecules are smooth, hard cylinders so that

$$E_{st}[R(s)] = 0 \text{ when } R(s) > 2a, \tag{2.17}$$

$$E_{st}[R(s)] = \infty \text{ when } R(s) \leq 2a. \tag{2.18}$$

This may be quite reasonable; we expect the effect of any steric chiral interaction arising from modelling the molecules as rough helices to be small, if the chiral effect from the finite ranged helix specific forces is strong. In addition, we find, for the range of parameters that we investigate using the KL theory, that the effective interaction from undulations with steric confinement is quite small. Most of the confinement comes from the electrostatic forces. The main role of steric repulsion, here, is to provide a cut-off that prevents one from overestimating the enhancement of the electrostatic contribution from undulations; which a smooth cylinder model should indeed be sufficient in providing.

**2.3 Approximating steric effects**

In principle, using the energy functional given by Eqs. (2.6)- (2.8), (2.10), (2.11), (2.17) and (2.18), we could construct the partition function. However, in practice, to make any analytical progress with the path integration is difficult with the steric term included. Instead, to estimate the effects of undulations, we adapt a simpler approach used in Refs. [24,13,9]. The steric confinement of a WLC molecule can be estimated quite well by constructing a harmonic pseudo potential that tries to reproduce steric confinement [24]. Following Refs. [9,13], we also estimate both a maximum and minimum cut-offs for fluctuations in $R(s)$ within the braid, $d_{\min}$ and $d_{\max}$. There indeed should be a maximum displacement for braid undulations, as pulling the braid apart at one location causes tightening of the braid at another location due to the elastic response and the braid geometry (see Fig 2).

To see how we introduce these cut-offs, let us first start by defining $r(s) = R(s) - R_0$, where $\langle R(s) \rangle = R_0$ is the mean braid diameter. For $r(s) > d_{max}$ we replace $r(s)$ in both the elastic and electrostatic energies with $d_{max}$, and for $r(s) < d_{min}$ we replace $r(s)$ with $d_{min}$, but we leave all derivatives of $r(s)$ untouched. It is this procedure that prevents an unphysical overestimation (or underestimation) of the average bending and electrostatic energies when we replace the steric interaction term with the harmonic potential [24]. The true steric interaction potential never allows for values of elastic bending energy and electrostatic energy, without these cut-offs, where $r(s) < d_{min}$ or $r(s) > d_{max}$.

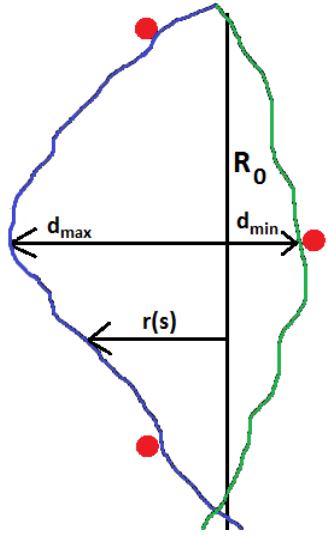

Fig 2. Schematic picture of confinement of a molecule by the other molecule in the braid. Here the red dots represent the confining molecule. The vertical spacing between dots is the super helical pitch $P$. The blue line represents a trajectory of the molecule that results in $d_{max}$, the maximum distance in $r(s)$. The green line represents a trajectory that results in $d_{min}$, the minimum distance in $r(s)$. The straight vertical black line represents the mean position of the molecule at $R(s) = R_0$.

Following the above procedure, we can write down an approximate form for the total energy functional

$$E_T[R(s), \eta(s), \Delta\Phi(s)] \approx \tilde{E}_{elast}[R(s), \eta(s), \Delta\Phi(s)] + \tilde{E}_{int}[R(s), \eta(s), \Delta\Phi(s)] + \tilde{E}_{st}[r(s)], \qquad (2.19)$$

where for the steric contribution we now have the harmonic pseudo potential

$$\tilde{E}_{st}[r(s)] = \int_{-\infty}^{\infty} \frac{k_0}{2} (R(s) - R_0)^2 = \int_{-\infty}^{\infty} \frac{k_0}{2} r(s)^2. \qquad (2.20)$$

For the elastic energy contribution we write

$$\tilde{E}_{elast}[R(s),\eta(s),\Delta\Phi(s)] \approx k_B T \int_0^{L_B} ds \Big( \mathcal{E}_{elast}\left(\Delta\Phi'(s), R''(s), R'(s), R_0 + d_{min}, \eta'(s), \eta(s)\right)$$
$$\theta(d_{min} - r(s)) + \mathcal{E}_{elast}\left(\Delta\Phi'(s), R''(s), R'(s), R_0 + r(s), \eta'(s), \eta(s)\right)\theta(d_{max} - r(s))\theta(r(s) - d_{min})$$
$$+ \mathcal{E}_{elast}\left(\Delta\Phi'(s), R''(s), R'(s), R_0 + d_{max}, \eta'(s), \eta(s)\right)\theta(r(s) - d_{max}) \Big).$$

(2.21)

For the interaction energy we may write

$$\tilde{E}_{int}[R(s),\eta(s),\Delta\Phi(s)] \approx k_B T \int_0^{L_B} ds \Big( \mathcal{E}_{int}\left(\Delta\Phi(s), R_0 + d_{min}, \eta(s)\right)\theta(d_{min} - r(s)) +$$
$$\mathcal{E}_{int}\left(\Delta\Phi(s), R_0 + r(s), \eta(s)\right)\theta(r(s) - d_{min})\theta(d_{max} - r(s)) + \mathcal{E}_{int}\left(\Delta\Phi(s), R_0 + d_{max}, \eta(s)\right)\theta(r(s) - d_{max}) \Big).$$

(2.22)

For $d_{min}$ we may simply choose $2a - R_0$, estimating $d_{max}$ is a little trickier for a braid. In the braid one can consider one of the molecules wrapping around to form a cage around the other (Fig 2) and vice versa. The most simplest assumption is to assume that this cage can be approximated by hard walled cylinder, and so setting $r_{max} \approx R - 2a$. We think, however, that this overestimates the degree of maximum mutual confinement of the molecules within the braid. Instead, we suppose that there is a maximum deflection length $\lambda_{max}$, the distance a molecule fluctuate away $R_0$ before returning back, that is of the order of the super-helical pitch, $P$, i.e.

$$\lambda_{max} \sim P = \pi R_0 / \tan(\eta_0 / 2), \tag{2.23}$$

where $\eta_0 = \langle \eta(s) \rangle$. This seems reasonable, as to be confined within a braid, the molecule must return back into the confining cage of the other molecule over the distance of one super helical pitch. To relate this to an a estimate for $d_{max}$, we then use the classic scaling formula [11]

$$\lambda_B = \left(\sqrt{2}d^2 l_p^b\right)^{1/3} \tag{2.24}$$

that relates the standard deflection length $\lambda_B$ to the root mean fluctuation amplitude $d$ (see Refs [9] and [11]). Therefore, combining both Eq. (2.23) and Eq. (2.24) with $\lambda_{max} = \lambda_B$, we obtain the estimate

$$d_{max} \approx \left(\frac{\pi R_0}{\tan \alpha_0}\right)^{3/2} \frac{1}{\left(l_p^b \sqrt{2}\right)^{1/2}}. \tag{2.25}$$

We find that for all the cases examined in the results section that in all of Eqs. (2.19)-(2.22) we can effectively set $d_{max}$ to infinity and not care too much about the accuracy of Eq. (2.25), when using Eqs. (2.12)-(2.16). This is because the estimated value of $d_{max}$ from Eq. (2.25) is sufficiently large for its difference from $d_{max} = \infty$ to be tiny in the average bending and electrostatic energy, which are used in the calculation. Another important difference from Ref [9] comes in how we estimate the effective spring constant of the harmonic term (Eq. (2.20)). We suppose that the mean squared displacement, when considering only steric interactions, is determined by an average of $d_{max}$ and $-d_{min}$, so that in terms of path integration over $r(s)$ we write

$$\left\langle r(s)^2 \right\rangle_{str} = \frac{\int Dr(s) r(s)^2 \exp[-E_{str}[r(s)]]}{\int Dr(s) \exp[-E_{str}[r(s)]]} \approx \left(d_{max} - d_{min}\right)^2 / 2, \quad (2.26)$$

where we use the energy functional

$$E_{str}[r(s)] = \frac{1}{2} \int_{L/2}^{-L/2} ds \left[ \frac{l_p^b}{2} \left( \frac{d^2 r(s)}{ds^2} \right)^2 + k_0 r(s)^2 \right] \quad (2.27)$$

in the Boltzmann weight of Eq. (2.26). From Eqs. (2.26) and (2.27), we estimate the effective spring constant to be

$$k_0 \approx \frac{2}{\left(d_{max} - d_{min}\right)^{8/3} \left(l_p^b\right)^{1/3}}. \quad (2.28)$$

**2.4 Variational calculation of the Free Energy**

Following Ref. [9], we then construct a variational principle where we build the following effective energy functional

$$E_{eff}[r(s), \delta\eta(s), \delta\Phi(s)] = k_B T \left( \int_0^{L_B} ds \left[ \frac{l_p^b}{4} \left( \frac{d^2 r(s)}{ds^2} \right)^2 + \frac{k_r r(s)^2}{2} \right] + \int_0^{L_B} ds \left[ \frac{l_p^b}{4} \left( \frac{d\delta\eta(s)}{ds} \right)^2 + \frac{k_\alpha \delta\eta(s)^2}{2} \right] \right.$$
$$\left. + k_B T \int_0^{L_B} ds \left[ \frac{l_p^h}{4} \left( \frac{d\delta\Phi(s)}{ds} \right)^2 + \frac{k_\Phi \delta\Phi(s)^2}{2} \right] \right),$$

$$(2.29)$$

where $\Delta\Phi(s) = \Delta\Phi_0(s) + \delta\Phi(s)$ and $\eta(s) = \eta_0 + \delta\eta(s)$; $\delta\Phi(s)$, $\delta\eta(s)$ and $r(s)$ are the thermal fluctuations about the mean fields $\Delta\Phi_0(s)$, $\eta_0$ and $R_0$, respectively. For $\sigma_H = 0$ $\Delta\Phi_0(s) = \Delta\bar{\Phi}$, a constant value, otherwise it depends on $s$ due to the random mismatch in helices $\Delta\Omega(s)$. In Appendix B of the Supplemental Material, for $\sigma_H = 1$, we consider a variational form for $\Delta\Phi_0(s)$ which is the same as that considered in Ref. [1], where both $\eta_0$ and $R_0$ are assumed constant with respect to $s$. We then calculate the variational Free Energy [9]

$$F_T = -kT \ln Z_{eff} + \left\langle \left\langle E_T[R(s),\eta(s),\Delta\Phi(s)] - E_{eff}[r(s),\delta\eta(s),\delta\Phi(s)] \right\rangle_0 \right\rangle_{\Delta\Omega}, \quad (2.30)$$

where

$$Z_{eff} = \int D\delta\Phi(s) \int D\delta\eta(s) \int Dr(s) \exp\left(-\frac{E_{eff}[r(s),\delta\eta(s),\delta\Phi(s)]}{k_B T}\right) \quad (2.31)$$

and the thermal average is given by

$$\left\langle E_T - E_{eff} \right\rangle_0 = \frac{1}{Z_{eff}} \int D\delta\Phi(s) \int D\delta\eta(s) \int Dr(s) \left(E_T - E_{eff}\right) \exp\left(-\frac{E_{eff}[r(s),\delta\eta(s),\delta\Phi(s)]}{k_B T}\right).$$

(2.32)

The averaging can be performed (see Appendix B of Supplemental Material) yielding the following energy function (where Eqs. (2.12)-(2.16) have been used)

$$\frac{F_T}{Lk_B T} = \frac{F_c}{Lk_B T} + \frac{l_p^b}{R_0^2} f(R_0, d_r; a) \left[\frac{3}{2} - 2\cos\eta_0 \exp\left(-\frac{\lambda_\eta}{2l_p^b}\right) + \frac{\cos 2\eta_0}{2} \exp\left(-\frac{2\lambda_\eta}{l_p^b}\right)\right]$$

$$+ \frac{2l_B(1-\theta)^2}{l_c^2} \frac{g_0(\kappa_D R_0, \kappa_D d_r, (R_0-2a)/d_r)}{(\kappa_D a K_1(\kappa_D a))^2} - \frac{4l_B}{l_c^2} \frac{[\zeta_1(f_1,f_2,\theta)]^2}{(\kappa_1 a K_1'(\kappa_1 a))^2} \cos\Delta\bar\Phi \exp\left(-\frac{\lambda_h^*}{2\lambda_c}\right)$$

$$\times \left[g_0(\kappa_1 R_0, \kappa_1 d_r, (R_0-2a)/d_r) + \bar{g}/\kappa_1 \sin\eta_0 \exp\left(-\frac{\lambda_\eta}{2l_p^b}\right) g_1(\kappa_1 R_0, \kappa_1 d_r, (R_0-2a)/d_r)\right]$$

$$+ \frac{4l_B}{l_c^2} \frac{[\zeta_2(f_1,f_2,\theta)]^2}{(\kappa_2 a K_2'(\kappa_2 a))^2} \cos 2\Delta\bar\Phi \exp\left(-\frac{2\lambda_h^*}{\lambda_c}\right)$$

$$\times \left[g_0(\kappa_2 R_0, \kappa_2 d_r, (R_0-2a)/d_r) + 4\bar{g}/\kappa_2 \sin\eta_0 \exp\left(-\frac{\lambda_\eta}{2l_p^b}\right) g_1(\kappa_2 R_0, \kappa_2 d_r, (R_0-2a)/d_r)\right]$$

$$+ \frac{2l_B}{l_c^2} \sum_{n=-\infty}^{\infty} \frac{g_{img}(n, \kappa_n R_0, \kappa_n d_r, (R_0-2a)/d_r; a)}{(\kappa_n a K_n'(\kappa_n a))^2} \left[\zeta_n(f_1,f_2,\theta)\right]^2,$$

(2.33)

and

$$\frac{F_c}{Lk_BT} = \frac{3}{2^{8/3} d_r^{2/3} \left(l_p^b\right)^{1/3}} + \frac{1}{4\lambda_\eta} + \frac{(l_p^h + \lambda_c)^2}{16\lambda_h \lambda_c l_p^h} + \frac{d_r^2}{(d_{max} - d_{min})^{8/3} \left(l_p^b\right)^{1/3}}, \qquad (2.34)$$

where

$$f(R_0, d_r; a) \simeq \frac{R_0}{d_r \sqrt{2\pi}} \int_{\left(\frac{2a}{R_0}-1\right)}^{\infty} dx \frac{1}{(1+x)^2} \exp\left(-\frac{x^2}{2}\left(\frac{R_0}{d_r}\right)^2\right) + \frac{1}{2}\frac{R_0^2}{(2a)^2}\left[1 - \mathrm{erf}\left(\frac{1}{\sqrt{2}} \frac{R_0 - 2a}{d_r}\right)\right], \qquad (2.35)$$

$$g_j(\kappa_n R_0, \kappa_D d_r, (R_0 - 2a)/d_r)$$
$$= \frac{1}{\sqrt{2\pi}} \int_{\frac{(2a-R_0)}{d_r}}^{\infty} dy K_j(\kappa_n R_0 + y\kappa_n d_r) \exp\left(-\frac{y^2}{2}\right) + \frac{K_j(2\kappa_n a)}{2}\left[1 - \mathrm{erf}\left(\frac{1}{\sqrt{2}} \frac{R_0 - 2a}{d_r}\right)\right], \qquad (2.36)$$

$$g_{img}(n, \kappa_n R_0, \kappa_n d_r, (R_0 - 2a)/d_r; a)$$
$$= -\sum_{j=-\infty}^{\infty} \int_{\frac{(2a-R_0)}{d_r}}^{\infty} dy K_{n-j}(\kappa_n R_0 + y\kappa_n d_r) K_{n-j}(\kappa_n R_0 + y\kappa_n d_r) \frac{I'_j(\kappa_n a)}{K'_j(\kappa_n a)} \exp\left(-\frac{y^2}{2}\right) \qquad (2.37)$$
$$-\frac{1}{2}\sum_{j=-\infty}^{\infty} K_{n-j}(2\kappa_n a) K_{n-j}(2\kappa_n a) \frac{I'_j(\kappa_n a)}{K'_j(\kappa_n a)}\left[1 - \mathrm{erf}\left(\frac{1}{\sqrt{2}} \frac{R_0 - 2a}{d_r}\right)\right].$$

In any other helix specific theory that has exponential decay of the helical harmonics of the interaction force with $R(s)$, the form of Eq. (2.36) may still be justified, although the pre-factors appearing in Eq. (2.33) may not be from the strict point of view of electrostatics. Here, we have eliminated $k_\Phi$ and $k_\eta$ in favour of the adaptation lengths $\lambda_h^* = 1/2\left(l_p^h/(2k_\Phi)\right)^{1/2}\left(1 + \lambda_c/l_p^h\right)$, $\lambda_\eta = \left(l_p^b/(2k_\eta)\right)^{1/2}$ and $k_r$ in terms of the mean square fluctuation amplitude $d_r = \sqrt{\langle r(s)^2 \rangle}$ which is related to $k_r$ through

$$k_r = \frac{1}{d_r^{8/3} 2^{5/3} \left(l_p^b\right)^{1/3}}. \qquad (2.38)$$

We find that $\left\langle \frac{l_p^b \sin\eta(s)}{2(R_0 + r(s))}\left(\frac{dR(s)}{ds}\right)\left(\frac{d\eta(s)}{ds}\right)\right\rangle_0 = 0$ and we have neglected

$\left\langle \frac{l_p^b(1-\cos\eta(s))}{2(R_0 + d_{max})^2}\left(\frac{dR(s)}{ds}\right)^2\right\rangle_0$. Which is reasonable, provided that

$$\left\langle \left(\frac{dR(s)}{ds}\right)^2 \right\rangle_0 \ll \frac{2\left(\frac{3}{2} - 2\cos\eta_0 \exp\left(-\frac{\lambda_\eta}{2l_p^b}\right) + \frac{1}{2}\cos 2\eta_0 \exp\left(-\frac{2\lambda_\eta}{l_p^b}\right)\right)}{1 - \cos\eta_0 \exp\left(-\frac{\lambda_\eta}{2l_p^b}\right)}, \quad (2.39)$$

which is indeed satisfied in the calculations of the results section. In Appendix C of the Supplemental Material we show the equations on $d_r$, $\lambda_h$, $\lambda_\eta$, $R_0$, $\Delta\Phi$, $\eta_0$ that arise from minimization of this free energy with respect to these variational parameters.

## 3. Results

Here, we investigate the effects of undulations on the results of Ref [1] for DNA. However, there is one additional difference; instead of the small angle formula for both the chiral torque and bending energy, we use trigonometric functions. We find that this only makes a slight difference to the results without undulations. The advantages of using such functions are that it makes thermal averaging easier and may provide a more rapidly convergent series in powers of $\sin\eta(s)$ when higher order corrections are considered. Nevertheless, for comparison purposes, as in Ref [1], we choose $f_1 = 0.4$ and $f_2 = 0.6$ and Debye length $\kappa_D^{-1} = 7\text{Å}$.

**3.1 Free energy**

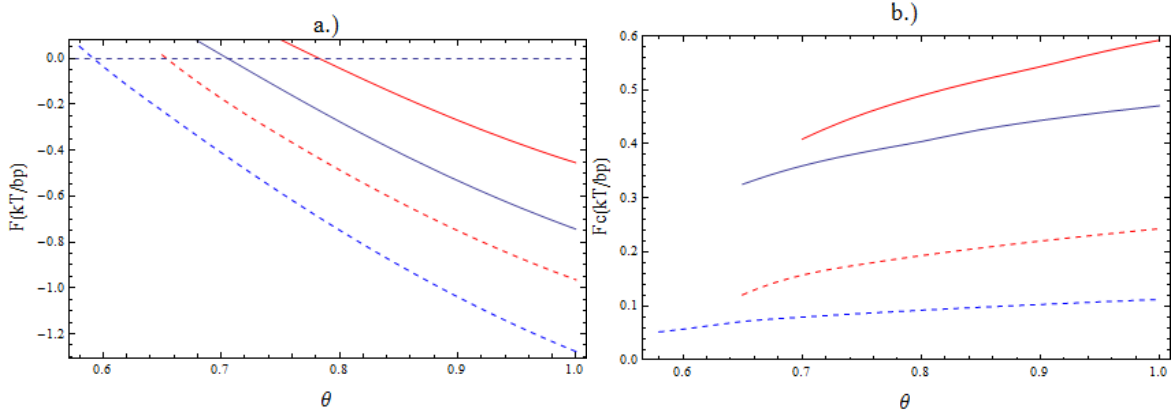

Fig.3. Braid confinement entropy increases threshold attraction for spontaneous braid formation. We plot a.) The total pairing free energy and b.) The confinement Free energy (See Eqs. (2.34) and (3.1)). Both are plotted as functions of the overall charge compensation $\theta$ with $f_1 = 0.4$, $f_2 = 0.6$ and $\kappa_D^{-1} = 7\text{Å}$. The blue curves are for a pair of Homologous molecules and the red curves are for Non-Homologous Molecules. The dashed lines are for the case without undulations, as in Ref [1], the solid lines are with undulations. We see that the confinement entropy is much larger when undulations are considered the drives up the value of $\theta$ where the braid becomes stable.

In Fig 3a we plot the total braiding free energy for homologous and non-homologous DNA molecules with and without undulations as a function of the charge compensation parameter $\theta$. This Free energy is Eq. (2.33) minimized with respect $d_r$, $\lambda_\eta$, $\lambda_h^*$, $R_0$, $\eta_0$ and $\Delta\bar{\Phi}$. For the braid to be stable $F<0$, otherwise any solution corresponds to a metastable state and the unpaired state of two isolated molecules (which has $F=0$) wins out. We see that the main effect of undulations is to push up the threshold value of $\theta$, at which the paired and the unpaired states have the same energy; we call this value $\theta_c$. For homologous molecules, including undulations pushes $\theta_c$ from $\approx 0.6$ to $\approx 0.7$, and for non-homologous molecules from $\theta_c \approx 0.65$ to $\theta_c \approx 0.75$. What is mainly responsible for this shift in $\theta_c$ is the contribution to the free energy $F_c$ due to the reduction in entropy when molecules are electrostatically confined within braided conformation. To see this we have plotted in Fig3b this contribution for the various cases considered in Fig3a. Without undulations [1], in Eq. (2.33) we set $d=0$, $\lambda_\eta = 0$ and $F_c$ is given by

$$\frac{F_c}{Lk_BT} = \frac{(l_p^h + \lambda_c)^2}{16\lambda_h \lambda_c l_p^h}. \tag{3.1}$$

In both Eqs. (2.34) and (3.1) we use the optimised values for $\lambda_h$, $d$ and $\lambda_\alpha$ (see Fig 4).

### 3.2 Fluctuation parameters

Looking at Fig. 4 we see the following trends. As we increase $\theta$, the parameters $\lambda_h$, $d$ and $\lambda_\eta$ all decrease, which means that the fluctuations in $\Delta\Phi(s)$, $R(s)$ and $\eta(s)$ all diminish. These trends drive the increasing entropy loss due to confinement with increasing $\theta$, resulting in the increase of free energy seen in Fig 3b. Also, both $d$ and $\lambda_\eta$ are quite small, which accounts for this large contribution to the free energy (Fig3b). The former is due to the strong electrostatic interactions, while the latter is due to rather large bending rigidly in $\eta(s)$. In Fig. 4a we see that undulations do not change much the value of $\lambda_h$; they only slightly increase $\lambda_h$ and this increase gets smaller as $\theta$ gets larger. This trend is completely consistent with both $d$ and $\lambda_\eta$ both decreasing with increasing $\theta$, as $\lambda_h$ without undulations is determined with both $d=0$ and $\lambda_\eta = 0$. In all cases, non-homologous molecules have the largest values of $\lambda_h / \lambda_c$, $d$ and $\lambda_\eta$ at fixed $\theta$; this is because they have the largest degree of fluctuations due intrinsic structural helix disorder, supressing helix specific electrostatic interactions.

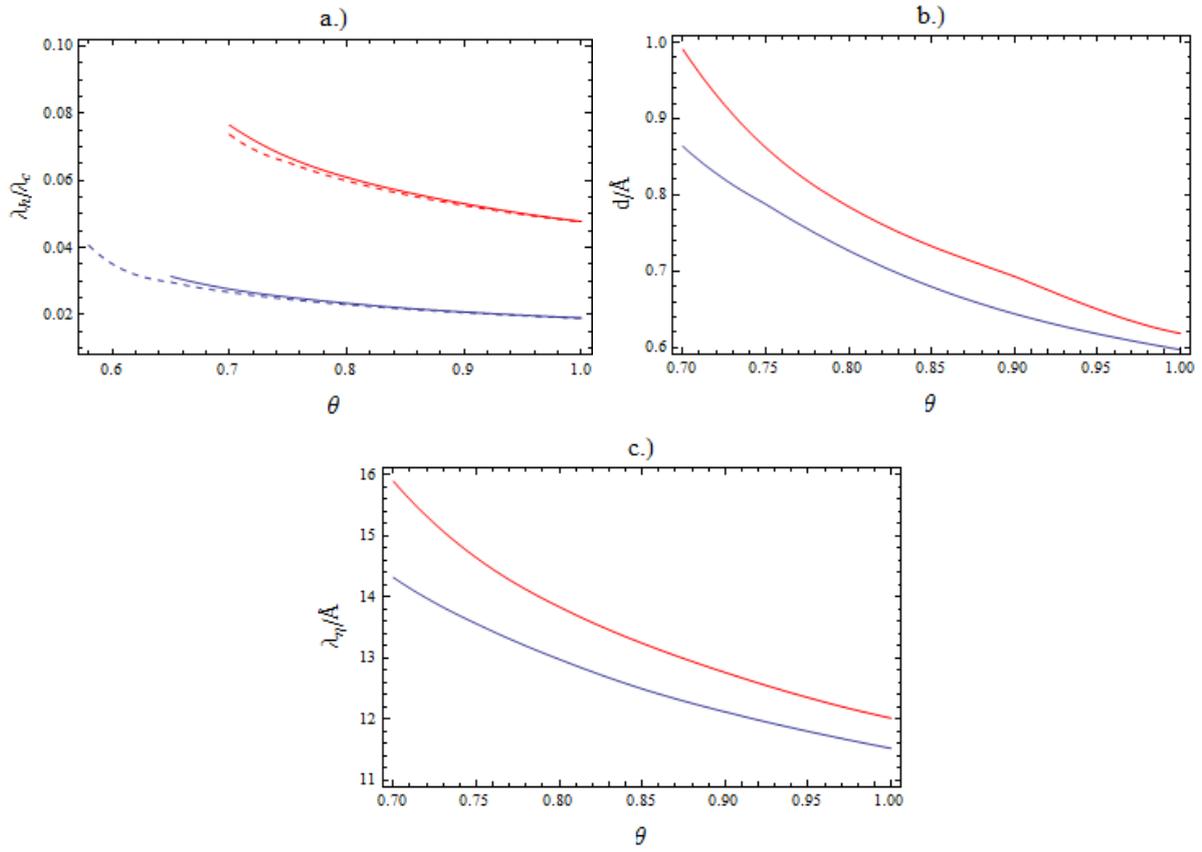

Fig 4. The braid fluctuation parameters. We show plots of a.) $\lambda_h$, the helical adaptation length b.) $d$ and c.) $\lambda_\eta$. Again, the values $f_1 = 0.4$, $f_2 = 0.6$ and $\kappa_D^{-1} = 7\text{Å}$ are used. Again, the blue curves are for a pair of Homologous molecules and the Red curves are for Non-Homologous Molecules. The dashed lines are for the case without undulations, as in Ref [1], the solid lines are with undulations. Non homologous molecules fluctuate more than homologous ones. $d$ is small due to electrostatic confinement and $\lambda_\eta$ due to bending elasticity.

### 3.3 Mean structural parameters

Now, let us examine the mean structural parameters. We find that $\overline{\Phi} \approx \pi/2$; it only very slightly increases as $\theta$ increases, and changes very little with undulations. Last of all, in Fig 5, we show $R_0$, $\eta_0$ and $P$, the mean braid diameter, tilt angle and super-coil pitch, respectively. As $\theta$ increases, $R_0$ decreases and $\eta_0$ increases which thereby causes a decrease in the supercoiling pitch of the braid. The reason why $\eta_0$ increases is simple; as we increase the value of $\theta$ while keeping $f_1 = 0.4$ and $f_2 = 0.6$ fixed, we make the variation in the surface charge density from positive to negative values more pronounced. This means that there is a stronger impetus for the molecules to want to be tilted with respect to each other, creating a stronger chiral torque (the size of the electrostatic terms in Eq. (2.11) that multiply $\sin \eta(s)$ increases). This increased torque pushes the equilibrium value of $\eta_0$ to larger values. The reduction in $R_0$ can be again be explained by the increase in electrostatic zipper attraction [25,6], when $\theta$ increases, which overcomes the image charge repulsion and steric forces more and pushes the two molecules closer together.

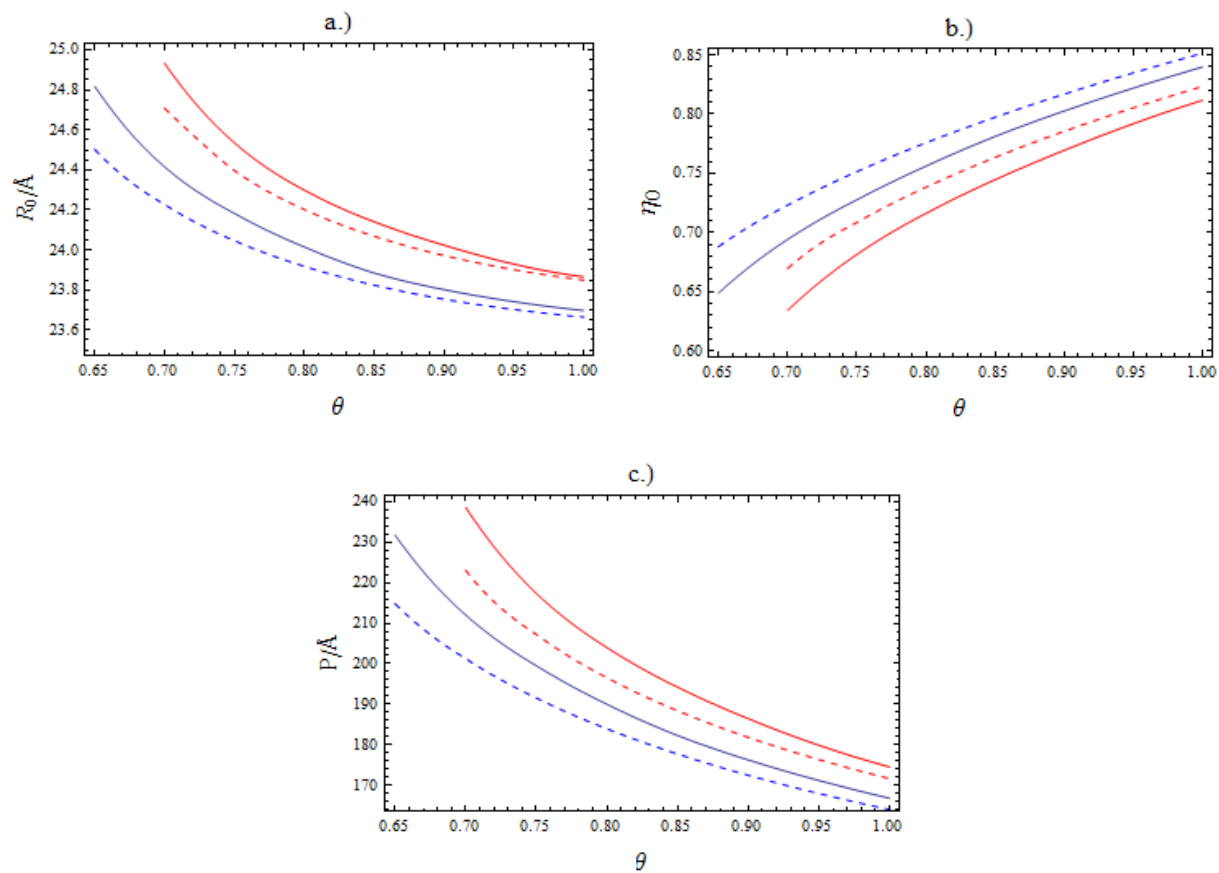

Fig 5. Braid structural parameters. We show plots of a.) $R_0$, the braid diameter b.) $\eta_0$, the tilt angle and c.) $P$, the braid supercoil pitch. Here we use the same values of the parameters and use of colours, solid and dashed lines as in the previous two figures. Undulations change $R_0$ and $\eta_0$ only very slightly. The balance of forces is not affected that much.

In Fig 5, we see that the effect of undulations is to slightly push both $R_0$ and $P$ to slightly larger values and $\eta_0$ to slightly smaller values. The balance of forces is not changed that much by undulations, this is reflected in the rather low values of $d$ and $\lambda_\eta$. The increase in $R_0$ can be explained by the fact that undulations cause the repulsive image charge terms in the electrostatic energy to be enhanced more than attractive terms arising from the direct electrostatic interaction. It is this increase in the repulsion from when $d=0$ (no undulations) that pushes up $R_0$ slightly. The values of $\eta_0$ are smaller presumably because fluctuations in $\eta$ weaken the chiral torque through $\lambda_\eta$, as well as the equilibrium $R_0$ being pushed to out to slightly larger values. Last of all, we observe that for homologous molecules, as opposed to non-homologous molecules, $R_0$ and $P$ are smaller and $\eta_0$ is larger. The value of $R_0$ is smaller for homologous molecules as the attractive helix specific forces are stronger, because the degree of large scale helical disorder $\lambda_h / \lambda_c$ is smaller than that for non-homologous molecules (see Fig 4a). The larger values of $\eta_0$ for homologous molecules is attributable again to the smaller value of $\lambda_h / \lambda_c$, which increases the strength of the chiral torque.

## 4. Discussion

In this paper, we have developed a theory to describe undulations in a braid where there are helix dependent forces. This theory should be able to be applied to charged helical macromolecules when their helical structure affects the interaction forces between them. We compared the results of the theory with those obtained in [1] for DNA like charge distributions using results from the electrostatic KL theory. It was found that the effect of the undulations was to increase considerably the amount of attraction needed for a braided structure to have lower free energy than two separated molecules, due to the loss of entropy from confinement. Braiding or partial braiding for homologous DNAs may still occur for two free DNA molecules in bulk monovalent salt solution; however, this state is likely to be meta-stable one. If this is so, the majority of DNA will be indeed unpaired and there will be only a few braided pairs. Therefore, it is reasonable to speculate that it may be a meta-stable state that is formed in the pairing experiments of [26]. This seems to be supported by a private communication [27] that suggested that majority of DNAs in those experiments maybe are, in fact, unpaired. This meta-stable state might be stabilized in sodium chloride solution due to molecular crowding effects and a reduction in configurational entropy due attachment of molecules to a substrate, or perhaps more subtler effects due to the capillary surface, so that the entropy cost for confinement of the DNA to a braid is lower.

For plectonemes, a simple model was presented to describe chiral effects in DNA closed loops [28]. In light of the results presented here, the model presented in Ref. [28] will have to be refined to take account of undulations, as well as helix non-ideality that was not included here. In such a system a tightly wound plectoneme state competes with a loosely wound state, for the latter chiral effects are not important. Including the confinement entropy and undulations is likely to favour this loosely wound state. This suggests that the asymmetry in the energy between left and right handed super-coils is over estimated without including undulations and non-ideality. Also, recent work [29] to do with single molecule twisting experiments [30] that uses the undulation theory of Ref [3] suggests that the chiral torque, if it is present for DNA, is too strong and part of this problem might be due to not self consistently estimating the confinement entropy. For a proper experimental test of whether helix specific forces are present- and if they are, their magnitude- undulations of the molecules within the braid certainly need to be estimated self-consistently. Therefore, we hope in a future work to modify the theory presented in this paper to closed loop super-coils and braids under mechanical forces [30,31] to obtain a more complete theory. Also, at a later stage, we will want to include undulations of the braid axis to complete the theory. However, one should point out, that to describe right handed DNA plectonemes properly, one may also have to take account of the B to Z transition, especially for GC sequences [32,33].

Of course, equilibrium separations $R_0 \approx 24\text{Å}$ are too close for the theory of [5], based on a bulk dielectric response, to be quantitatively valid. One of the hopes behind this study was that undulations would push up the value of $R_0$ considerably, but this effect seems to be slight, the confinement entropy matters much more. Therefore, instead of Eqs. (2.12)-(2.16), perhaps a better approach would be to obtain empirical forms for $E_{img}(R)$, $E_0^{(n)}(R)$ and $E_1^{(n)}(R)$ from simulation data with realistic groove binding potentials and structure of counter-ions. Indeed, the calculations presented in this paper could then be redone to reflect any such refinement in the interaction energy; if helical structure does matter at all.

It may also be theoretically interesting and insightful to redo these calculations for interaction potentials in the strong coupling limit [7], where correlations between ions are strong. Here, instead of an electrostatic zipper, one has a correlation zipper [8]. Though the effects of helical adaptation will be more or less described by the same theory as described here, the azimuthal alignment will be of the opposite sense to that of the electrostatic zipper [7,8]; the phosphates of both molecules want to come close to each other instead of trying to be far apart as possible, so as to generate the strongest correlation forces. It would be interesting to see what effect this has on the formation and structure of the braid, as in certain cases correlation effects could be very important.

## Acknowledgements

D.J. Lee would like to acknowledge useful discussions with A. A. Korynshev and R. Cortini. He would also like to acknowledge the support of the Human Frontiers Science Program (grant RGP0049/2010-C102). This work has also been inspired by joint work that has been supported by the United Kingdom Engineering and Physical Sciences Research Council (grant EP/H004319/1).

# Supplemental Material

## Appendix A: Derivation of the Energy Functional

### The Elastic Energy Term

The bending energy is given by

$$E_b = k_B T \int_{-L/2}^{L/2} ds \left[ \frac{l_p^b}{2} \left( \frac{d\hat{\mathbf{t}}_1(s)}{ds} \right)^2 + \frac{l_p^b}{2} \left( \frac{d\hat{\mathbf{t}}_2(s)}{ds} \right)^2 \right]. \tag{A.1}$$

Our task is to evaluate both $\frac{d\hat{\mathbf{t}}_1(s)}{ds}$ and $\frac{d\hat{\mathbf{t}}_2(s)}{ds}$. From Eq. (2.1) of the main text we find that

$$\hat{\mathbf{t}}_1(s) = \frac{dz(s)}{ds}\hat{\mathbf{k}} - \frac{1}{2}\frac{dR(s)}{ds}\hat{\mathbf{d}}(s) - \frac{R(s)}{2}\frac{d\hat{\mathbf{d}}(s)}{ds}, \tag{A.2}$$

$$\hat{\mathbf{t}}_2(s) = \frac{dz(s)}{ds}\hat{\mathbf{k}} + \frac{1}{2}\frac{dR(s)}{ds}\hat{\mathbf{d}}(s) + \frac{R(s)}{2}\frac{d\hat{\mathbf{d}}(s)}{ds}. \tag{A.3}$$

We can represent $\hat{\mathbf{d}}(s)$ as

$$\hat{\mathbf{d}}(s) = \cos\theta(s)\hat{\mathbf{i}} + \sin\theta(s)\hat{\mathbf{j}} \tag{A.4}$$

as it is perpendicular to $\hat{\mathbf{k}}$ along which the braid axis lies. This allows us to write Eqs. (A.2) and (A.3) as

$$\hat{\mathbf{t}}_1(s) = \frac{dz(s)}{ds}\hat{\mathbf{k}} - \left( \frac{1}{2}\frac{dR(s)}{ds}\cos\theta(s) - \frac{R(s)}{2}\frac{d\theta(s)}{ds}\sin\theta(s) \right)\hat{\mathbf{i}} - \left( \frac{1}{2}\frac{dR(s)}{ds}\sin\theta(s) + \frac{R(s)}{2}\frac{d\theta(s)}{ds}\cos\theta(s) \right)\hat{\mathbf{j}}, \tag{A.5}$$

$$\hat{\mathbf{t}}_2(s) = \frac{dz(s)}{ds}\hat{\mathbf{k}} + \left( \frac{1}{2}\frac{dR(s)}{ds}\cos\theta(s) - \frac{R(s)}{2}\frac{d\theta(s)}{ds}\sin\theta(s) \right)\hat{\mathbf{i}} + \left( \frac{1}{2}\frac{dR(s)}{ds}\sin\theta(s) + \frac{R(s)}{2}\frac{d\theta(s)}{ds}\cos\theta(s) \right)\hat{\mathbf{j}}. \tag{A.6}$$

The requirement that both $\hat{\mathbf{t}}_1(s)$ and $\hat{\mathbf{t}}_2(s)$ are unitary and that $\hat{\mathbf{t}}_1(s).\hat{\mathbf{t}}_2(s) = \cos\eta(s)$ requires that we write

$$\left( \frac{dz}{ds} \right)^2 = 1 - \frac{1}{4}\left( \frac{dR(s)}{ds} \right)^2 - \frac{R(s)^2}{4}\left( \frac{d\theta(s)}{ds} \right)^2, \tag{A.7}$$

$$\cos\eta(s) = \left( \frac{dz}{ds} \right)^2 - \frac{1}{4}\left( \frac{dR(s)}{ds} \right)^2 - \frac{R(s)^2}{4}\left( \frac{d\theta(s)}{ds} \right)^2. \tag{A.8}$$

Combining both Eqs. (A.7) and (A.8) we may write

$$1 - \frac{1}{2}\left(\frac{dR(s)}{ds}\right)^2 - \frac{R(s)^2}{2}\left(\frac{d\theta(s)}{ds}\right)^2 = \cos\eta(s). \tag{A.9}$$

If we suppose that $\frac{1}{2}\left(\frac{dR(s)}{ds}\right)^2 \ll \cos\eta(s)$, we can approximate from Eqs. (A.7) and (A.9)

$$\frac{R(s)}{2}\frac{d\theta(s)}{ds} \simeq -\sin\left(\frac{\eta(s)}{2}\right), \qquad \left(\frac{dz}{ds}\right) \simeq \cos\left(\frac{\eta(s)}{2}\right). \tag{A.10}$$

Therefore, we in this approximation we obtain

$$\hat{\mathbf{t}}_1(s) \simeq \cos\left(\frac{\eta(s)}{2}\right)\hat{\mathbf{k}} - \left(\frac{1}{2}\frac{dR(s)}{ds}\cos\theta(s) + \sin\left(\frac{\eta(s)}{2}\right)\sin\theta(s)\right)\hat{\mathbf{i}} - \left(\frac{1}{2}\frac{dR(s)}{ds}\sin\theta(s) - \sin\left(\frac{\eta(s)}{2}\right)\cos\theta(s)\right)\hat{\mathbf{j}},$$

(A.11)

$$\hat{\mathbf{t}}_2(s) \simeq \cos\left(\frac{\eta(s)}{2}\right)\hat{\mathbf{k}} + \left(\frac{1}{2}\frac{dR(s)}{ds}\cos\theta(s) + \sin\left(\frac{\eta(s)}{2}\right)\sin\theta(s)\right)\hat{\mathbf{i}} + \left(\frac{1}{2}\frac{dR(s)}{ds}\sin\theta(s) - \sin\left(\frac{\eta(s)}{2}\right)\cos\theta(s)\right)\hat{\mathbf{j}}.$$

(A.12)

Differentiation of the tangent vectors then yields following

$$\frac{d\hat{\mathbf{t}}_1(s)}{ds} = -\frac{1}{2}\frac{d\eta(s)}{ds}\sin\left(\frac{\eta(s)}{2}\right)\hat{\mathbf{k}}$$
$$-\left(\left(\frac{1}{2}\frac{d^2R(s)}{ds^2} - \frac{2}{R(s)}\sin^2\left(\frac{\eta(s)}{2}\right)\right)\cos\theta(s) + \left(\frac{dR(s)}{ds}\frac{1}{R(s)}\sin\left(\frac{\eta(s)}{2}\right) + \frac{1}{2}\frac{d\eta(s)}{ds}\cos\left(\frac{\eta(s)}{2}\right)\right)\sin\theta(s)\right)\hat{\mathbf{i}}$$
$$+\left(\left(\frac{dR(s)}{ds}\frac{1}{R(s)}\sin\left(\frac{\eta(s)}{2}\right) + \frac{1}{2}\frac{d\eta(s)}{ds}\cos\left(\frac{\eta(s)}{2}\right)\right)\cos\theta(s) - \left(\frac{1}{2}\frac{d^2R(s)}{ds^2} - \frac{2}{R(s)}\sin^2\left(\frac{\eta(s)}{2}\right)\right)\sin\theta(s)\right)\hat{\mathbf{j}},$$

(A.13)

$$\frac{d\hat{\mathbf{t}}_2(s)}{ds} = -\frac{1}{2}\frac{d\eta(s)}{ds}\sin\left(\frac{\eta(s)}{2}\right)\hat{\mathbf{k}}$$
$$+\left(\left(\frac{1}{2}\frac{d^2R(s)}{ds^2} - \frac{2}{R(s)}\sin^2\left(\frac{\eta(s)}{2}\right)\right)\cos\theta(s) + \left(\frac{dR(s)}{ds}\frac{1}{R(s)}\sin\left(\frac{\eta(s)}{2}\right) + \frac{1}{2}\frac{d\eta(s)}{ds}\cos\left(\frac{\eta(s)}{2}\right)\right)\sin\theta(s)\right)\hat{\mathbf{i}}$$
$$-\left(\left(\frac{dR(s)}{ds}\frac{1}{R(s)}\sin\left(\frac{\eta(s)}{2}\right) + \frac{1}{2}\frac{d\eta(s)}{ds}\cos\left(\frac{\eta(s)}{2}\right)\right)\cos\theta(s) - \left(\frac{1}{2}\frac{d^2R(s)}{ds^2} - \frac{2}{R(s)}\sin^2\left(\frac{\eta(s)}{2}\right)\right)\sin\theta(s)\right)\hat{\mathbf{j}}.$$

(A.14)

Substituting Eqs. (A.13) and (A.14) into the elastic energy,(A.1) we obtain

$$E_b = k_B T \int_{-L/2}^{L/2} ds\, l_p^b \left[ \frac{1}{4}\left(\frac{d^2 R(s)}{ds^2}\right)^2 - \frac{1}{R(s)}(1-\cos\eta(s))\left(\frac{d^2 R(s)}{ds^2}\right) + \frac{(1-\cos\eta(s))^2}{R(s)^2} \right.$$

$$\left. + \left(\frac{dR(s)}{ds}\right)^2 \frac{(1-\cos\eta(s))}{2R(s)^2} + \frac{1}{2}\frac{d\eta(s)}{ds}\left(\frac{dR(s)}{ds}\right)\frac{1}{R(s)}\sin(\eta(s)) + \frac{1}{4}\left(\frac{d\eta(s)}{ds}\right)^2 \right].$$

(A.15)

By parts integration of the second term in Eq. (A.15) we obtain

$$E_b = k_B T \int_{-L/2}^{L/2} ds\, l_p^b \left[ \frac{1}{4}\left(\frac{d^2 R(s)}{ds^2}\right)^2 + \frac{(1-\cos\eta(s))^2}{R(s)^2} - \left(\frac{dR(s)}{ds}\right)^2 \frac{(1-\cos\eta(s))}{2R(s)^2} \right.$$

$$\left. + \frac{3}{2}\frac{d\eta(s)}{ds}\left(\frac{dR(s)}{ds}\right)\frac{1}{R(s)}\sin(\eta(s)) + \frac{1}{4}\left(\frac{d\eta(s)}{ds}\right)^2 \right].$$

(A.16)

The elastic energy of helical distortions can be written as [9]

$$E_h = \int_{-L/2}^{L/2} ds\, \frac{l_p^h}{2}\left[\left(g_1(s) - g_1^0(s)\right)^2 + \left(g_2(s) - g_2^0(s)\right)^2\right].$$

(A.17)

For straight molecules the local twist densities $g_1(s)$ and $g_2(s)$ are given by

$$g_1(s) = \frac{d\Phi_1(s)}{ds} + \bar{g}, \quad g_2(s) = \frac{d\Phi_2(s)}{ds} + \bar{g},$$

(A.18)

where $\Phi_1$ and $\Phi_2$ are the helical phases defined in Ref [9] and $\bar{g} = 2\pi/H$, the parameter $H$ is the average pitch of the helical molecule. For DNA, the local twist densities of the unstressed states are given by

$$g_1^0(s) = \frac{\Omega_1^0(s) - \bar{g}h_1^0(s)}{\langle h \rangle} + \bar{g}, \quad g_2^0(s) = \frac{\Omega_2^0(s) - \bar{g}h_2^0(s)}{\langle h \rangle} + \bar{g}.$$

(A.19)

Here $\Omega_1^0(s)$ and $\Omega_2^0(s)$ are the pattern of twist angles between each adjacent base pair and $h_1^0(s)$ and $h_2^0(s)$ are the patterns of rises, for molecules 1 and 2 in their relaxed states, respectively. These patterns depend on the sequence of base pairs. For molecules that is form a perfect helix in its ground state $g_1^0(s) = g_2^0(s) = \bar{g}$.

For a molecular braid the twist densities $g_1(s)$ and $g_2(s)$ can be computed through the general formulas:

$$g_1(s) = \hat{\mathbf{u}}_1(s) \cdot \frac{d\hat{\mathbf{v}}_1(s)}{ds}, \quad g_2(s) = \hat{\mathbf{u}}_2(s) \cdot \frac{d\hat{\mathbf{v}}_2(s)}{ds},$$

(A.20)

where $\hat{\mathbf{u}}_1(s) = \hat{\mathbf{t}}_1(s) \times \hat{\mathbf{v}}_1(s)$ and $\hat{\mathbf{u}}_2(s) = \hat{\mathbf{t}}_2(s) \times \hat{\mathbf{v}}_2(s)$. We first may write, using Eq. (2.5) of the main text,

$$\frac{d\hat{\mathbf{v}}_1(s)}{ds} = \frac{d\phi_1(s)}{ds}\left(-\sin\phi_1(s)\hat{\mathbf{d}}_1(s) + \cos\phi_1(s)\hat{\mathbf{n}}_1(s)\right) + \cos\phi_1(s)\frac{d\hat{\mathbf{d}}_1(s)}{ds} + \sin\phi_1(s)\frac{d\hat{\mathbf{n}}_1(s)}{ds},$$

(A.21)

$$\frac{d\hat{\mathbf{v}}_2(s)}{ds} = \frac{d\phi_2(s)}{ds}\left(-\sin\phi_2(s)\hat{\mathbf{d}}_2(s) + \cos\phi_2(s)\hat{\mathbf{n}}_2(s)\right) + \cos\phi_2(s)\frac{d\hat{\mathbf{d}}_2(s)}{ds} + \sin\phi_2(s)\frac{d\hat{\mathbf{n}}_2(s)}{ds}.$$

(A.22)

We can also write

$$\hat{\mathbf{u}}_1(s) = \left(-\sin\phi_1(s)\hat{\mathbf{d}}_1(s) + \cos\phi_1(s)\hat{\mathbf{n}}_1(s)\right), \quad \hat{\mathbf{u}}_2(s) = \left(-\sin\phi_2(s)\hat{\mathbf{d}}_2(s) + \cos\phi_2(s)\hat{\mathbf{n}}_2(s)\right). \quad \text{(A.23)}$$

Therefore, from Eqs. (A.20)-(A.23) we obtain

$$g_1(s) = \frac{d\phi_1(s)}{ds} - \hat{\mathbf{d}}_1(s).\frac{d\hat{\mathbf{n}}_1(s)}{ds}, \qquad g_2(s) = \frac{d\phi_2(s)}{ds} - \hat{\mathbf{d}}_2(s).\frac{d\hat{\mathbf{n}}_2(s)}{ds}, \qquad \text{(A.24)}$$

where we have utilized

$$\hat{\mathbf{d}}_1(s).\frac{d\hat{\mathbf{d}}_1(s)}{ds} = 0, \ \hat{\mathbf{d}}_2(s).\frac{d\hat{\mathbf{d}}_2(s)}{ds} = 0, \ \hat{\mathbf{n}}_1(s).\frac{d\hat{\mathbf{n}}_1(s)}{ds} = 0, \ \hat{\mathbf{n}}_2(s).\frac{d\hat{\mathbf{n}}_2(s)}{ds} = 0,$$

$$\hat{\mathbf{d}}_1(s).\frac{d\hat{\mathbf{n}}_1(s)}{ds} = -\hat{\mathbf{n}}_1(s).\frac{d\hat{\mathbf{d}}_1(s)}{ds}, \ \hat{\mathbf{d}}_2(s).\frac{d\hat{\mathbf{n}}_2(s)}{ds} = -\hat{\mathbf{n}}_2(s).\frac{d\hat{\mathbf{d}}_2(s)}{ds}. \quad \text{(A.25)}$$

All of these relations in Eq. (A.25) come from the fact that
$\hat{\mathbf{d}}_1(s).\hat{\mathbf{d}}_1(s) = \hat{\mathbf{d}}_2(s).\hat{\mathbf{d}}_2(s) = \hat{\mathbf{n}}_1(s).\hat{\mathbf{n}}_1(s) = \hat{\mathbf{n}}_2(s).\hat{\mathbf{n}}_2(s) = 1$ and $\hat{\mathbf{n}}_1(s).\hat{\mathbf{d}}_1(s) = \hat{\mathbf{n}}_2(s).\hat{\mathbf{d}}_2(s) = 0$.

Next we need to compute $\hat{\mathbf{n}}_1(s)$, $\hat{\mathbf{d}}_1(s)$, $\hat{\mathbf{n}}_2(s)$, $\hat{\mathbf{d}}_2(s)$ using Eqs. (A.4), (A.5) and (A.6), as well as Eqs. (2.2) and (2.3) of the main text, we find

$$\hat{\mathbf{n}}_1(s) = \left(1 - \frac{1}{4}\left(\frac{dR(s)}{ds}\right)^2\right)^{-1/2}\left(-\sin\theta(s)\frac{dz(s)}{ds}\hat{\mathbf{i}} + \cos\theta(s)\frac{dz(s)}{ds}\hat{\mathbf{j}} + \frac{R(s)}{2}\frac{d\theta(s)}{ds}\hat{\mathbf{k}}\right), \quad \text{(A.26)}$$

$$\hat{\mathbf{n}}_2(s) = \left(1 - \frac{1}{4}\left(\frac{dR(s)}{ds}\right)^2\right)^{-1/2}\left(-\sin\theta(s)\frac{dz(s)}{ds}\hat{\mathbf{i}} + \cos\theta(s)\frac{dz(s)}{ds}\hat{\mathbf{j}} - \frac{R(s)}{2}\frac{d\theta(s)}{ds}\hat{\mathbf{k}}\right), \quad \text{(A.27)}$$

$$\hat{\mathbf{d}}_1(s) = \left(1 - \frac{1}{4}\left(\frac{dR(s)}{ds}\right)^2\right)^{-1/2} \left\{\cos\theta(s)\left[\left(\frac{dz(s)}{ds}\right)^2 + \frac{R(s)^2}{4}\left(\frac{d\theta(s)}{ds}\right)^2\right] + \frac{R(s)}{4}\frac{dR(s)}{ds}\frac{d\theta(s)}{ds}\sin\theta(s)\right\}\hat{\mathbf{i}}$$
$$+ \left\{\sin\theta(s)\left[\left(\frac{dz(s)}{ds}\right)^2 + \frac{R(s)^2}{4}\left(\frac{d\theta(s)}{ds}\right)^2\right] - \frac{R(s)}{4}\frac{dR(s)}{ds}\frac{d\theta(s)}{ds}\cos\theta(s)\right\}\hat{\mathbf{j}} + \frac{1}{2}\frac{dz(s)}{ds}\frac{dR(s)}{ds}\hat{\mathbf{k}},$$

(A.28)

$$\hat{\mathbf{d}}_2(s) = \left(1 - \frac{1}{4}\left(\frac{dR(s)}{ds}\right)^2\right)^{-1/2} \left\{\cos\theta(s)\left[\left(\frac{dz(s)}{ds}\right)^2 + \frac{R(s)^2}{4}\left(\frac{d\theta(s)}{ds}\right)^2\right] + \frac{R(s)}{4}\frac{dR(s)}{ds}\frac{d\theta(s)}{ds}\sin\theta(s)\right\}\hat{\mathbf{i}}$$
$$+ \left\{\sin\theta(s)\left[\left(\frac{dz(s)}{ds}\right)^2 + \frac{R(s)^2}{4}\left(\frac{d\theta(s)}{ds}\right)^2\right] - \frac{R(s)}{4}\frac{dR(s)}{ds}\frac{d\theta(s)}{ds}\cos\theta(s)\right\}\hat{\mathbf{j}} - \frac{1}{2}\frac{dz(s)}{ds}\frac{dR(s)}{ds}\hat{\mathbf{k}}.$$

(A.29)

If we suppose that $\frac{1}{2}\left(\frac{dR(s)}{ds}\right)^2 \ll \cos\eta(s)$ we can use Eq. (A.10) to write

$$\hat{\mathbf{n}}_1(s) \simeq -\sin\theta(s)\cos\left(\frac{\eta(s)}{2}\right)\hat{\mathbf{i}} + \cos\theta(s)\cos\left(\frac{\eta(s)}{2}\right)\hat{\mathbf{j}} - \sin\left(\frac{\eta(s)}{2}\right)\hat{\mathbf{k}},$$  (A.30)

$$\hat{\mathbf{n}}_2(s) \simeq -\sin\theta(s)\cos\left(\frac{\eta(s)}{2}\right)\hat{\mathbf{i}} + \cos\theta(s)\cos\left(\frac{\eta(s)}{2}\right)\hat{\mathbf{j}} + \sin\left(\frac{\eta(s)}{2}\right)\hat{\mathbf{k}},$$  (A.31)

$$\hat{\mathbf{d}}_1(s) \simeq \left\{\cos\theta(s) - \frac{1}{2}\frac{dR(s)}{ds}\sin\left(\frac{\eta(s)}{2}\right)\sin\theta(s)\right\}\hat{\mathbf{i}}$$
$$+ \left\{\sin\theta(s) + \frac{1}{2}\frac{dR(s)}{ds}\sin\left(\frac{\eta(s)}{2}\right)\cos\theta(s)\right\}\hat{\mathbf{j}} + \frac{1}{2}\cos\left(\frac{\eta(s)}{2}\right)\frac{dR(s)}{ds}\hat{\mathbf{k}},$$

(A.32)

$$\hat{\mathbf{d}}_2(s) \simeq \left\{\cos\theta(s) - \frac{1}{2}\frac{dR(s)}{ds}\sin\left(\frac{\eta(s)}{2}\right)\sin\theta(s)\right\}\hat{\mathbf{i}}$$
$$+ \left\{\sin\theta(s) + \frac{1}{2}\frac{dR(s)}{ds}\sin\left(\frac{\eta(s)}{2}\right)\cos\theta(s)\right\}\hat{\mathbf{j}} - \frac{1}{2}\cos\left(\frac{\eta(s)}{2}\right)\frac{dR(s)}{ds}\hat{\mathbf{k}}.$$

(A.33)

We then compute the derivatives

$$\frac{d\hat{\mathbf{n}}_1(s)}{ds} = \left(\frac{1}{2}\frac{d\eta(s)}{ds}\sin\left(\frac{\eta(s)}{2}\right)\sin\theta(s) + \frac{\cos\theta(s)\sin\eta(s)}{R(s)}\right)\hat{\mathbf{i}}$$
$$- \left(\frac{1}{2}\frac{d\eta(s)}{ds}\sin\left(\frac{\eta(s)}{2}\right)\cos\theta(s) - \frac{\sin\theta(s)\sin\eta(s)}{R(s)}\right)\hat{\mathbf{j}} - \frac{1}{2}\frac{d\eta(s)}{ds}\cos\left(\frac{\eta(s)}{2}\right)\hat{\mathbf{k}},$$

(A.34)

$$\frac{d\hat{\mathbf{n}}_2(s)}{ds} = \left(\frac{1}{2}\frac{d\eta(s)}{ds}\sin\left(\frac{\eta(s)}{2}\right)\sin\theta(s) + \frac{\cos\theta(s)\sin\eta(s)}{R(s)}\right)\hat{\mathbf{i}}$$
$$-\left(\frac{1}{2}\frac{d\eta(s)}{ds}\sin\left(\frac{\eta(s)}{2}\right)\cos\theta(s) - \frac{\sin\theta(s)\sin\eta(s)}{R(s)}\right)\hat{\mathbf{j}} + \frac{1}{2}\frac{d\eta(s)}{ds}\cos\left(\frac{\eta(s)}{2}\right)\hat{\mathbf{k}}$$

(A.35)

Therefore, we obtain from Eqs. (A.24) and (A.32)-(A.35) the following

$$g_1(s) = \frac{d\phi_1(s)}{ds} - \frac{\sin\eta(s)}{R} + \frac{1}{4}\frac{dR(s)}{ds}\frac{d\eta(s)}{ds} \quad g_2(s) = \frac{d\phi_2(s)}{ds} - \frac{\sin\eta(s)}{R} + \frac{1}{4}\frac{dR(s)}{ds}\frac{d\eta(s)}{ds} \quad (A.36)$$

Then, using Eq. (A.35), we rewrite Eq. (A.17) as

$$E_{tw} = \int_{-L/2}^{L/2} ds \frac{l_c^h}{2}\left[\left(\frac{d\phi_1(s)}{ds} - \frac{\sin\eta(s)}{R} + \frac{1}{4}\frac{d\eta(s)}{ds}\frac{dR(s)}{ds} - g_1^0(s)\right)^2\right.$$
$$+ \left.\left(\frac{d\phi_2(s)}{ds} - \frac{\sin\eta(s)}{R} + \frac{1}{4}\frac{d\eta(s)}{ds}\frac{dR(s)}{ds} - g_2^0(s)\right)^2\right]$$
$$= \int_{-L/2}^{L/2} ds \frac{l_c^h}{4}\left(\frac{d\Delta\Phi(s)}{ds} - \frac{\Delta\Omega(s)}{h}\right)^2 + \frac{l_c^h}{4}\left(\frac{d\bar{\phi}(s)}{ds} - \frac{2\sin\eta(s)}{R} + \frac{1}{2}\frac{d\eta(s)}{ds}\frac{dR(s)}{ds} - \left(g_1^0(s) + g_2^0(s)\right)\right)^2,$$

(A.37)

where $\Delta\Omega(s) = h(g_1^0(s) - g_2^0(s))$, $\Delta\Phi(s) = \phi_1(s) - \phi_2(s)$ and $\bar{\phi}(s) = \phi_1(s) + \phi_2(s)$.

### The decoupling of the $\bar{\phi}$ degrees of freedom

Provided that $\frac{dR(s)}{ds} \ll 1$, from the result given in [21] for a symmetric braid we can write the electrostatic energy as

$$E_{els}[R(s),\eta(s),\Delta\Phi(s),g_1(s),g_2(s)] \approx k_B T \int_0^{L_B} ds \left(\frac{E_{img}(R(s),g_1(s))}{2} + \frac{E_{img}(R(s),g_2(s))}{2}\right.$$
$$+ \left.\sum_{n=0}^{2}\left\{E_0^{(n)}\left(R(s),\frac{1}{2}(g_1(s)+g_2(s))\right) + \sin\eta(s)E_1^{(n)}\left(R(s),\frac{1}{2}(g_1(s)+g_2(s))\right)\right\}\cos n\Delta\Phi(s)\right),$$

(A.38)

with

$$E_{img}(R,g) = -\frac{2l_B}{l_c^2}\sum_{n=-\infty}^{\infty}\sum_{j=-\infty}^{\infty}\frac{K_{n-j}(\kappa_n(g)R)K_{n-j}(\kappa_n(g)R)}{(\kappa_n(g)aK_n'(\kappa_n(g)a))^2}\frac{I_j'(\kappa_n(g)a)}{K_j'(\kappa_n(g)a)}\zeta_n^2, \quad (A.39)$$

$$E_0^{(0)}(R,g) = \frac{2l_B(1-\theta)^2}{l_c^2}\frac{K_0(\kappa_D R)}{(\kappa_D a K_1(\kappa_D a))^2}, \quad (A.40)$$

$$E_0^{(n)}(R,g) = \frac{4l_B}{l_c^2} \frac{(-1)^n K_0(\kappa_n(g)R)\zeta_n^2}{(\kappa_n(g)aK_n'(\kappa_n(g)a))^2} \quad \text{for } n \neq 0, \tag{A.41}$$

$$E_1^{(n)}(R,g) = \frac{4l_B n^2 ag}{l_c^2} \frac{(-1)^n K_1(\kappa_n(g)R)\zeta_n^2}{(K_n'(\kappa_n(g)a))^2 (\kappa_n(g)a)^3}, \tag{A.42}$$

Where $\kappa_n(g) = \sqrt{\kappa_D^2 + n^2 g^2}$ and for DNA like charge distributions we have

$$\zeta_n(f_1, f_2, \theta) = \theta\left[f_1 + (-1)^n f_2\right] - \cos n\tilde{\phi}_s, \tag{A.43}$$

(see the main text for an explanation of various parameters contained in Eq. (A.43)). This result (Eqs. (A.38)-(A.43)) should be valid provided that $\kappa_D l_p^h, \kappa_D l_p^b \gg 1$ so that $\frac{d\eta(s)}{ds}$ and higher order derivatives can be neglected from the electrostatic calculation.

Using the definitions for the $g_j(s)$s contained in Eq. (A.36) we can rewrite Eq. (A.38) as

$$\begin{aligned}E_{els}[R(s),\eta(s),\Delta\Phi(s),\bar{\phi}(s)] &\approx k_B T \int_0^{L_B} ds \Bigg( E_{img}\left(R(s), \frac{1}{2}\left(\frac{d\Delta\Phi(s)}{ds} + \frac{d\bar{\phi}(s)}{ds}\right) - \frac{\sin\eta(s)}{R} + \frac{1}{4}\frac{dR(s)}{ds}\frac{d\eta(s)}{ds}\right) \\ &+ E_{img}\left(R(s), \frac{1}{2}\left(\frac{d\bar{\phi}(s)}{ds} - \frac{d\Delta\Phi(s)}{ds}\right) - \frac{\sin\eta(s)}{R} + \frac{1}{4}\frac{dR(s)}{ds}\frac{d\eta(s)}{ds}\right) \\ &+ \sum_{n=0}^{2}\Bigg\{E_0^{(n)}\left(R(s), \frac{d\bar{\phi}(s)}{ds} - \frac{\sin\eta(s)}{R} + \frac{1}{4}\frac{dR(s)}{ds}\frac{d\eta(s)}{ds}\right) \\ &+ \sin\eta(s) E_1^{(n)}\left(R(s), \frac{d\bar{\phi}(s)}{ds} - \frac{\sin\eta(s)}{R} + \frac{1}{4}\frac{dR(s)}{ds}\frac{d\eta(s)}{ds}\right)\Bigg\}\cos n\Delta\Phi(s)\Bigg).\end{aligned}$$

(A.44)

Now provided that $\frac{d\Delta\Phi(s)}{ds} \ll \frac{d\bar{\phi}}{ds} - \frac{2\sin\eta(s)}{R} + \frac{1}{4}\frac{dR(s)}{ds}\frac{d\eta(s)}{ds}$ we can write Eq. (A.44) as

$$\begin{aligned}E_{els}[R(s),\eta(s),\Delta\Phi(s),g(s)] &\approx k_B T \int_0^{L_B} ds \Big(E_{img}\left(R(s), g(s)\right) \\ &+ \sum_{n=0}^{2}\left\{E_0^{(n)}\left(R(s), g(s)\right) + \sin\eta(s) E_1^{(n)}\left(R(s), g(s)\right)\right\}\cos n\Delta\Phi(s)\Big),\end{aligned} \tag{A.45}$$

where $g(s) = \frac{d\bar{\phi}(s)}{ds} - \frac{\sin\eta(s)}{R} + \frac{1}{4}\frac{d\eta(s)}{ds}\frac{dR(s)}{ds}$. Let us suppose that $g(s)$ lies close to $g_0(s) = (g_1^0(s) + g_2^0(s))/2$ and perform a Taylor expansion of Eq. (A.45) yielding

$$E_{els}[R(s),\eta(s),\Delta\Phi(s),g(s)] \approx$$

$$k_BT\int_0^{L_B} ds\left[\left(E_{img}(R(s),g_0(s))\right)+\sum_{n=0}^{2}\left\{E_0^{(n)}(R(s),g_0(s))+\sin\eta(s)E_1^{(n)}(R(s),g_0(s))\right\}\cos n\Delta\Phi(s)\right.$$

$$+\left(\frac{\partial E_{img}(R(s),g_0(s))}{\partial g_0(s)}+\sum_{n=0}^{2}\left\{\frac{\partial E_0^{(n)}(R(s),g_0(s))}{\partial g_0(s)}+\sin\eta(s)\frac{\partial E_1^{(n)}(R(s),g_0(s))}{\partial g_0(s)}\right\}\cos n\Delta\Phi(s)\right)(g(s)-g_0(s))$$

$$\left.+\left(\frac{\partial^2 E_{img}(R(s),g_0(s))}{\partial g_0(s)^2}+\sum_{n=0}^{2}\left\{\frac{\partial^2 E_0^{(n)}(R(s),g_0(s))}{\partial g_0(s)^2}+\sin\eta(s)\frac{\partial^2 E_1^{(n)}(R(s),g_0(s))}{\partial g_0(s)^2}\right\}\cos n\Delta\Phi(s)\right)\frac{(g(s)-g_0(s))^2}{2}\right].$$

(A.46)

Let us now consider the total energy

$$E_T = E_b + E_h + E_{els}.$$

(A.47)

Now provided that

$$\frac{k_BT}{2}\left(\frac{\partial^2 E_{img}(R(s),g_0(s))}{\partial g_0(s)^2}+\sum_{n=0}^{2}\left\{\frac{\partial^2 E_0^{(n)}(R(s),g_0(s))}{\partial g_0(s)^2}+\sin\eta(s)\frac{\partial^2 E_1^{(n)}(R(s),g_0(s))}{\partial g_0(s)^2}\right\}\cos n\Delta\Phi(s)\right) \ll l_p^h,$$

(A.48)

we have a minimum close to $g_0(s)$ which is

$$g(s) = g_0(s) - \frac{k_BT}{2l_p^h}\left(\frac{\partial E_{img}(R(s),)g_0(s)}{\partial g_0(s)}+\sum_{n=0}^{2}\left\{\frac{\partial E_0^{(n)}(R(s),g_0(s))}{\partial g_0(s)}+\sin\eta(s)\frac{\partial E_1^{(n)}(R(s),g_0(s))}{\partial g_0(s)}\right\}\cos n\Delta\Phi(s)\right).$$

(A.49)

Therefore provided that

$$\frac{k_BT}{2l_c^h}\left(\frac{\partial E_{img}(R(s),)g_0(s)}{\partial g_0(s)}+\sum_{n=0}^{2}\left\{\frac{\partial E_0^{(n)}(R(s),g_0(s))}{\partial g_0(s)}+\sin\eta(s)\frac{\partial E_1^{(n)}(R(s),g_0(s))}{\partial g_0(s)}\right\}\cos n\Delta\Phi(s)\right) \ll g_0(s),$$

(A.50)

as well as Eq. (A.48), we can approximately set $g(s)$ to $g_0(s)$. Furthermore, provided that $\kappa_n\lambda_c \gg 0$ the variation in $g_0(s)$ due to helix distortions is unimportant in the electrostatic energy and $g_0(s)$ can be replaced by $\bar{g}$ its mean value. Therefore $\bar{\phi}(s)$ decouple from the problem and we can then write $E_T$ in the form of Eq. (2.4) of the main text.

## Appendix B: Averaging terms in the Free energy

### *Derivation of general formulas for Gaussian averaging*

Suppose we want to calculate the average of a function

$$\left\langle F\left(X(s),\frac{dX(s)}{ds}\right)\right\rangle_X = \frac{\int DX(s)F\left(X(s),\frac{dX(s)}{ds}\right)\exp[-E[X(s)]]}{\int DX(s)\exp[-E[X(s)]]}, \quad (B.1)$$

where

$$E[X(s)] = \int_{-L/2}^{L/2} ds \int_{-L/2}^{L/2} ds' X(s) G^{-1}(s-s') X(s'). \quad (B.2)$$

To perform this average we can first start by writing

$$\left\langle F\left(X(s),\frac{dX(s)}{ds}\right)\right\rangle_X = \int_{-\infty}^{\infty} dx \int_{-\infty}^{\infty} dx' F(x,x')\left\langle \delta(x-X(s))\delta\left(x'-\frac{dX(s)}{ds}\right)\right\rangle_X, \quad (B.3)$$

which can be further rewritten as

$$\left\langle F\left(X(s),\frac{dX(s)}{ds}\right)\right\rangle_X = \int_{-\infty}^{\infty} dx \int_{-\infty}^{\infty} dx' \int_{-\infty}^{\infty} dp \int_{-\infty}^{\infty} dp' F(x,x')\exp(-ipx)\exp(-ip'x')\left\langle \exp(ipX(s))\exp\left(ip'\frac{dX(s)}{ds}\right)\right\rangle_X. \quad (B.4)$$

We can then write

$$\left\langle \exp(ipX(s))\exp\left(ip'\frac{dX(s)}{ds}\right)\right\rangle_X$$

$$= \frac{\int D\tilde{X}(k)\exp\left(\frac{i}{2\pi}\int_{-\infty}^{\infty}(p+ikp')\tilde{X}(k)\exp(ikx)\right)\exp\left(-\frac{1}{4\pi}\int_{-\infty}^{\infty}\tilde{X}(k)G^{-1}(k)\tilde{X}(-k)dk\right)}{\int D\tilde{X}(k)\exp\left(-\frac{1}{4\pi}\int_{-\infty}^{\infty}\tilde{X}(k)G^{-1}(k)\tilde{X}(-k)dk\right)}. \quad (B.5)$$

This can be rewritten as

$$\left\langle \exp(ipX(s))\exp\left(ip'\frac{dX(s)}{ds}\right)\right\rangle_X$$

$$= \frac{\int D\tilde{X}(k)\exp\left(-\frac{1}{4\pi}\int_{-\infty}^{\infty}\left(\tilde{X}(k)-G(k)(p+ikp')\exp(ikx)\right)G^{-1}(k)\left(\tilde{X}(-k)-G(k)(p-ikp')\exp(-ikx)\right)dk\right)}{\int D\tilde{X}(k)\exp\left(-\frac{1}{4\pi}\int_{-\infty}^{\infty}\tilde{X}(k)G^{-1}(k)\tilde{X}(-k)dk\right)}$$

$$\times \exp\left(\frac{1}{4\pi}\int_{-\infty}^{\infty} dk G(k)\left(p^2+k^2 p'^2\right)\right) = \exp\left(\frac{1}{4\pi}\int_{-\infty}^{\infty} dk G(k)\left(p^2+k^2 p'^2\right)\right).$$

$$(B.6)$$

Substitution of Eq. (B.6) into Eq.(B.4) yields the general formula

$$\left\langle F\left(X(s),\frac{dX(s)}{ds}\right)\right\rangle_X = \frac{1}{(2\pi)^2}\int_{-\infty}^{\infty}dx\int_{-\infty}^{\infty}dx'\int_{-\infty}^{\infty}dp\int_{-\infty}^{\infty}dp'F(x,x')\exp(ipx)\exp(ip'x')\exp\left(-\frac{p^2d_X^2}{2}\right)\exp\left(-\frac{p^2\theta_X^2}{2}\right)$$

$$=\frac{1}{2\pi d_X\theta_X}\int_{-\infty}^{\infty}dx\int_{-\infty}^{\infty}dx'F(x,x')\exp\left(-\frac{1}{2}\frac{x^2}{d_X^2}\right)\exp\left(-\frac{1}{2}\frac{x'^2}{\theta_X^2}\right),$$

(B.7)

where

$$d_X^2 = \frac{1}{2\pi}\int_{-\infty}^{\infty}dkG(k),\quad \theta_X^2 = \frac{1}{2\pi}\int_{-\infty}^{\infty}dkk^2G(k).$$

(B.8)

### *Averaging terms in the elastic energy*

We want to compute the Free energy (Eq. (2.30) of main text)

$$-kT\ln Z_{\text{eff}} + \left\langle\left\langle E_T[R(s),\eta(s),\Delta\Phi(s)] - E_{\text{eff}}[r(s),\delta\eta(s),\delta\Phi(s)]\right\rangle_0\right\rangle_{\Delta\Omega}$$

$$= -k_BT\ln Z_{\text{eff}} + \left\langle\left\langle\Delta\tilde{E}_{\text{elast}}[R(s),\eta(s),\Delta\Phi(s)]\right\rangle_0\right\rangle_{\Delta\Omega} + \left\langle\left\langle\Delta\tilde{E}_{el}[R(s),\eta(s),\Delta\Phi(s)]\right\rangle_0\right\rangle_{\Delta\Omega}$$

(B.9)

$$+\frac{k_BT}{2}\int_{-L/2}^{L/2}ds\left((k_0-k_r)\left\langle r(s)^2\right\rangle_0 - k_{\Delta\Phi}\left\langle\delta\Phi(s)^2\right\rangle_0 - k_\eta\left\langle\delta\eta(s)^2\right\rangle_0\right).$$

For the elastic energy contribution we need to perform the average

$$\left\langle\left\langle\Delta\tilde{E}_{\text{elast}}[R(s),\eta(s),\Delta\Phi(s)]\right\rangle_0\right\rangle_{\Delta\Omega}$$

$$= k_BT\int_{-L/2}^{L/2}ds\left[\left\langle\left\langle\Delta\mathcal{E}_h\left(\frac{d\Delta\Phi(s)}{ds}\right)\right\rangle_0\right\rangle_{\Delta\Omega} + l_p^b\left\langle\Delta\mathcal{E}_b\left(\frac{dr(s)}{ds},R_0+r(s),\frac{d\eta(s)}{ds},\eta(s)\right)\right\rangle_0\right],$$

(B.10)

where

$$\left\langle\Delta\mathcal{E}_b\left(\frac{dr(s)}{ds},r(s),\frac{d\eta(s)}{ds},\eta(s)\right)\right\rangle_0 = \left\langle U_b\left(\frac{dr(s)}{ds},R_0+d_{\min},\frac{d\eta(s)}{ds},\eta(s)\right)\theta(d_{\min}-r(s))\right\rangle_0$$

$$\left\langle U_b\left(\frac{dr(s)}{ds},R_0+r(s),\frac{d\eta(s)}{ds},\eta(s)\right)\theta(r(s)-d_{\min})\theta(d_{\max}-r(s))\right\rangle_0$$

$$+\left\langle U_b\left(\frac{dr(s)}{ds},R_0+d_{\max},\frac{d\eta(s)}{ds},\eta(s)\right)\theta(r(s)-d_{\max})\right\rangle_0,$$

(B.11)

and

$$U_b\left(\frac{dr(s)}{ds}, R_0 + r(s), \frac{d\eta(s)}{ds}, \eta(s)\right) =$$

$$\frac{(1-\cos\eta(s))^2}{(R_0+r(s))^2} - \frac{(1-\cos\eta(s))}{2(R_0+r(s))^2}\left(\frac{dr(s)}{ds}\right)^2 + \frac{3\sin\eta(s)}{2(R_0+r(s))^2}\left(\frac{dr(s)}{ds}\right)\left(\frac{d\eta(s)}{ds}\right),$$
(B.12)

also

$$\left\langle\left\langle \Delta\mathcal{E}_h\left(\frac{d\Delta\Phi(s)}{ds}\right)\right\rangle_0\right\rangle_{\Delta\Omega} = \left\langle\left\langle \frac{l_p^h}{4}\left(\frac{d\Delta\Phi(s)}{ds} - \sigma_H\Delta\Omega(s)\right)^2\right\rangle_0\right\rangle_{\Delta\Omega} - \left\langle\frac{l_p^h}{4}\left(\frac{d\delta\Phi(s)}{ds}\right)^2\right\rangle_{\Delta\Omega}$$
(B.13)

Let's first consider the contribution from bending terms. First of all (from Eqs. (B.11) and (B.12) combined) we first consider the average

$$\left\langle \frac{(1-\cos\eta(s))^2}{(R_0+d_{min})^2}\theta(d_{min}-r(s)) + \frac{(1-\cos\eta(s))^2}{(R_0+r(s))^2}\theta(r(s)-d_{min})\theta(d_{max}-r(s)) + \frac{(1-\cos\eta(s))^2}{(R_0+d_{max})^2}\theta(r(s)-d_{max})\right\rangle_0$$

$$= \left\langle \frac{3}{2} - 2\cos\eta(s) + \frac{\cos(2\eta(s))}{2}\right\rangle_\eta \left\langle \frac{\theta(d_{min}-r(s))}{(R_0+d_{min})^2} + \frac{\theta(r(s)-d_{min})\theta(d_{max}-r(s))}{(R_0+r(s))^2} + \frac{\theta(r(s)-d_{max})}{(R_0+d_{max})^2}\right\rangle_r.$$
(B.14)

Here, the subscripts on $\eta$ and $r$ on the averaging brackets denotes thermal averaging over $\delta\eta(s)$ and $r(s)$ respectively. Then, using the general averaging formula (Eq. (B.7)) we may write

$$\left\langle \left(\frac{3}{2}\right) - 2\cos\eta(s) + \frac{\cos(2\eta(s))}{2}\right\rangle_\eta = \frac{3}{2} - \frac{1}{\sqrt{2\pi}d_\eta}\int_{-\infty}^{\infty}d\eta'\left(2\cos(\eta_0+\eta') - \frac{\cos(2(\eta_0+\eta'))}{2}\right)\exp\left(-\frac{1}{2}\frac{\eta'^2}{d_\eta^2}\right)$$

$$= \frac{3}{2} - \frac{1}{\sqrt{2\pi}d_\eta}\int_{-\infty}^{\infty}d\eta'\left(2\cos\eta_0\cos\eta' - \frac{\cos(2\eta_0)\cos(2\eta')}{2}\right)\exp\left(-\frac{1}{2}\frac{\eta'^2}{d_\eta^2}\right)$$

$$= \frac{3}{2} - 2\cos\eta_0\exp\left(-\frac{d_\eta^2}{2}\right) + \frac{\cos(2\eta_0)}{2}\exp(-2d_\eta^2),$$
(B.15)

$$\left\langle \frac{\theta(d_{min}-r(s))}{(R_0+d_{min})^2} + \frac{\theta(r(s)-d_{min})\theta(d_{max}-r(s))}{(R_0+r(s))^2} + \frac{\theta(r(s)-d_{max})}{(R_0+d_{max})^2}\right\rangle_r$$

$$= \frac{1}{\sqrt{2\pi}d_r}\int_{-\infty}^{\infty}dr\left[\frac{\theta(d_{min}-r)}{(R_0+d_{min})^2} + \frac{\theta(r-d_{min})\theta(d_{max}-r)}{(R_0+r)^2} + \frac{\theta(r-d_{max})}{(R_0+d_{max})^2}\right]\exp\left(-\frac{r^2}{2d_r^2}\right)$$

$$= \frac{1}{2}\frac{\left(1-\operatorname{erf}\left(d_{min}/(d_r\sqrt{2})\right)\right)}{(R_0+d_{min})^2} + \frac{1}{2}\frac{\left(1-\operatorname{erf}\left(d_{max}/(d_r\sqrt{2})\right)\right)}{(R_0+d_{max})^2} + \frac{1}{\sqrt{2\pi}}\int_{-d_{min}/d_r}^{d_{max}/d_r}dr\frac{1}{(R_0+d_rr)^2}\exp\left(-\frac{r^2}{2}\right).$$
(B.16)

Next, in Eqs. (B.11) and (B.12) we consider

$$\left\langle \frac{(1-\cos\eta(s))}{2(R_0+d_{min})^2}\left(\frac{dr(s)}{ds}\right)^2 \theta(d_{min}-r(s)) + \frac{(1-\cos\eta(s))}{(R_0+r(s))^2}\left(\frac{dr(s)}{ds}\right)^2 \theta(r(s)-d_{min})\theta(d_{max}-r(s)) + \right.$$

$$\left. + \frac{(1-\cos\eta(s))}{(R_0+d_{max})^2}\left(\frac{dr(s)}{ds}\right)^2 \theta(r(s)-d_{max})\right\rangle_0 = \left\langle (1-\cos\eta(s))\right\rangle_\eta$$

$$\left\langle \frac{\theta(d_{min}-r(s))}{2(R_0+d_{min})^2}\left(\frac{dr(s)}{ds}\right)^2 + \frac{\theta(r(s)-d_{min})\theta(d_{max}-r(s))}{(R_0+r(s))^2}\left(\frac{dr(s)}{ds}\right)^2 + \frac{\theta(r(s)-d_{max})}{(R_0+d_{max})^2}\left(\frac{dr(s)}{ds}\right)^2 \right\rangle_r.$$

(B.17)

We find that

$$\left\langle (1-\cos\eta(s))\right\rangle_\eta = 1-\cos\eta_0 \exp\left(-\frac{d_r^2}{2}\right),$$
(B.18)

$$\left\langle \frac{\theta(d_{min}-r(s))}{2(R_0+d_{min})^2}\left(\frac{dr(s)}{ds}\right)^2 + \frac{\theta(r(s)-d_{min})\theta(d_{max}-r(s))}{(R_0+r(s))^2}\left(\frac{dr(s)}{ds}\right)^2 + \frac{\theta(r(s)-d_{max})}{(R_0+d_{max})^2}\left(\frac{dr(s)}{ds}\right)^2 \right\rangle_r$$

$$= \theta_r^2 \left[ \frac{1}{2}\frac{\left(1-\mathrm{erf}\left(d_{min}/(d_r\sqrt{2})\right)\right)}{(R_0+d_{min})^2} + \frac{1}{2}\frac{\left(1-\mathrm{erf}\left(d_{max}/(d_r\sqrt{2})\right)\right)}{(R_0+d_{max})^2} + \frac{1}{\sqrt{2\pi}}\int_{-d_{min}/d_r}^{d_{max}/d_r} dr \frac{1}{(R_0+d_r r)^2}\exp\left(-\frac{r^2}{2}\right) \right].$$

(B.19)

Provided that Eq. (2.39) of the main text is satisfied ($d_\eta = \lambda_\eta / l_P^b$ as is shown below), it is possible to neglect Eq. (B.19). Last of all we consider

$$\frac{3}{2}\left\langle \sin\eta(s)\left(\frac{dR(s)}{ds}\right)\left(\frac{d\eta(s)}{ds}\right)\left(\frac{\theta(d_{min}-r(s))}{(R_0+d_{min})} + \frac{\theta(r(s)-d_{min})\theta(d_{max}-r(s))}{(R_0+r(s))} + \frac{\theta(r(s)-d_{max})}{(R_0+d_{max})^2}\right)\right\rangle_0$$

$$= \frac{3}{2}\left\langle \sin\eta(s)\left(\frac{d\eta(s)}{ds}\right)\right\rangle_\eta \left\langle \left(\frac{dR(s)}{ds}\right)\left(\frac{\theta(d_{min}-r(s))}{(R_0+d_{min})} + \frac{\theta(r(s)-d_{min})\theta(d_{max}-r(s))}{(R_0+r(s))} + \frac{\theta(r(s)-d_{max})}{(R_0+d_{max})^2}\right)\right\rangle_r.$$

(B.20)

Through the general formula, Eq. (B.7), it is easy to show that Eq. (B.20) evaluates to zero. Now we consider the twisting contribution. By first writing $\Delta\Phi(s) = \Delta\Phi_0(s) + \delta\Phi(s)$ we can express Eq. (B.13) as

$$\left\langle \left\langle \Delta\mathcal{E}_h\left(\frac{d\Delta\Phi(s)}{ds}\right)\right\rangle_0\right\rangle_{\Delta\Omega} = \left\langle \frac{l_P^h}{4}\left(\frac{d\Delta\Phi_0(s)}{ds} - \sigma_H \Delta\Omega(s)\right)^2\right\rangle_{\Delta\Omega}.$$
(B.21)

As before in [1], to evaluate this ensemble average we choose a variational trial function for $\Delta\Phi_0(s)$ of the form

$$\Delta\Phi_0(s) = \Delta\bar{\Phi} + \frac{\sigma_H}{2\langle h \rangle} \int_{-L_B/2}^{L_B/2} \frac{\Delta\Omega(s')(s-s')}{|s-s'|} \exp\left(-\frac{|s-s'|}{\lambda_h}\right). \tag{B.22}$$

The average in (B.21) can be easily performed using Eq. (2.9) of the main text and $L_B$ taken to be infinite. This yields the result

$$\left\langle\left\langle \Delta\mathcal{E}_h\left(\frac{d\Delta\Phi(s)}{ds}\right)\right\rangle_0\right\rangle_{\Delta\Omega} = \frac{l_p^h \sigma_H}{8\lambda_c \lambda_h}. \tag{B.23}$$

### *Averaging terms in the electrostatic energy*

We can write from Eqs. (2.11), (2.22) of the main text

$$\left\langle\left\langle \tilde{E}_{el}[R(s), \eta(s), \Delta\Phi(s)]\right\rangle_0\right\rangle_{\Delta\Omega} \approx$$

$$k_B T \int_0^{L_B} ds \Big( \left\langle E_{img}(R_0 + r(s))\theta(r(s) - d_{min})\theta(d_{max} - r(s))\right\rangle_0$$

$$+ \left\langle E_{img}(R_0 + d_{min})\theta(d_{min} - r(s))\right\rangle_0 + \left\langle E_{img}(R_0 + d_{max})\theta(r(s) - d_{max})\right\rangle_0$$

$$+ \sum_{n=0}^{2}\Big[ \left\langle\left\langle \{E_0^{(n)}(R_0+r(s)) + \sin\eta(s)E_1^{(n)}(R_0+r(s))\}\theta(r(s)-d_{min})\theta(d_{max}-r(s))\cos(n\Delta\Phi(s))\right\rangle_0\right\rangle_{\Delta\Omega}$$

$$+ \left\langle\left\langle \{E_0^{(n)}(R_0+d_{min}) + \sin\eta(s)E_1^{(n)}(R_0+d_{min})\}\theta(d_{min}-r(s))\cos(n\Delta\Phi(s))\right\rangle_0\right\rangle_{\Delta\Omega}$$

$$+ \left\langle\left\langle \{E_0^{(n)}(R_0+d_{max}) + \sin\eta(s)E_1^{(n)}(R_0+d_{max})\}\theta(r(s)-d_{max})\cos(n\Delta\Phi(s))\right\rangle_0\right\rangle_{\Delta\Omega}\Big]\Big). \tag{B.24}$$

Using the form given by the KL theory for $E_{img}(R(s))$ (Eq. (2.12) of the main text) we can write

$$\left\langle E_{img}(R_0 + r(s))\theta(r(s) - d_{min})\theta(d_{max} - r(s))\right\rangle_0$$

$$+ \left\langle E_{img}(R_0 + d_{min})\theta(d_{min} - r(s))\right\rangle_0 + \left\langle E_{img}(R_0 + d_{max})\theta(r(s) - d_{max})\right\rangle_0$$

$$= -\frac{2l_B}{l_c^2} \sum_{n=-\infty}^{\infty} \sum_{j=-\infty}^{\infty} \frac{\zeta_n^2}{(\kappa_n a K_n'(\kappa_n a))^2} \frac{I_j'(\kappa_n a)}{K_j'(\kappa_n a)} \tag{B.25}$$

$$\times \Big[ \left\langle K_{n-j}(\kappa_n(R_0+r(s)))K_{n-j}(\kappa_n(R_0+r(s)))\theta(r(s)-d_{min})\theta(d_{max}-r(s))\right\rangle_r$$

$$+ K_{n-j}(\kappa_n(R_0+d_{min}))K_{n-j}(\kappa_n(R_0+d_{min}))\left\langle\theta(d_{min}-r(s))\right\rangle_r$$

$$+ K_{n-j}(\kappa_n(R_0+d_{max}))K_{n-j}(\kappa_n(R_0+d_{max}))\left\langle\theta(r(s)-d_{max})\right\rangle_r \Big].$$

Then, using Eq. (B.7), we can write

$$\langle E_{img}(R_0 + r(s))\theta(r(s) - d_{min})\theta(d_{max} - r(s))\rangle_0$$
$$+ \langle E_{img}(R_0 + d_{min})\theta(d_{min} - r(s))\rangle_0 + \langle E_{img}(R_0 + d_{max})\theta(r(s) - d_{max})\rangle_0$$
$$= -\frac{2l_B}{l_c^2} \sum_{n=-\infty}^{\infty} \sum_{j=-\infty}^{\infty} \frac{\zeta_n^2}{(\kappa_n a K_n'(\kappa_n a))^2} \frac{I_j'(\kappa_n a)}{K_j'(\kappa_n a)}$$
$$\times \left[ \frac{1}{d\sqrt{2\pi}} \int_{d_{min}}^{d_{max}} dr K_{n-j}(\kappa_n(R_0 + r)) K_{n-j}(\kappa_n(R_0 + r)) \exp\left(-\frac{r^2}{2d_r^2}\right) \right. \quad \text{(B.26)}$$
$$+ \frac{1}{2} K_{n-j}(\kappa_n(R_0 + d_{min})) K_{n-j}(\kappa_n(R_0 + d_{min})) \left(1 - \text{erf}\left(\frac{d_{min}}{d_r\sqrt{2}}\right)\right)$$
$$\left. + \frac{1}{2} K_{n-j}(\kappa_n(R_0 + d_{max})) K_{n-j}(\kappa_n(R_0 + d_{max})) \left(1 - \text{erf}\left(\frac{d_{max}}{d_r\sqrt{2}}\right)\right) \right].$$

Using the form of the effective energy functional, Eq. (2.32) of the main text, we are able to write

$$\sum_{n=0}^{2} \left[ \left\langle \left\langle \left\{ E_0^{(n)}(R_0 + r(s)) + \sin\eta(s) E_1^{(n)}(R_0 + r(s)) \right\} \theta(r(s) - d_{min})\theta(d_{max} - r(s)) \cos n\Delta\Phi(s) \right\rangle_0 \right\rangle_{\Delta\Omega} \right.$$
$$+ \left\langle \left\langle \left\{ E_0^{(n)}(R_0 + d_{min}) + \sin\eta(s) E_1^{(n)}(R_0 + d_{min}) \right\} \theta(d_{min} - r(s)) \cos n\Delta\Phi(s) \right\rangle_0 \right\rangle_{\Delta\Omega}$$
$$\left. + \left\langle \left\langle \left\{ E_0^{(n)}(R_0 + d_{max}) + \sin\eta(s) E_1^{(n)}(R_0 + d_{max}) \right\} \theta(r(s) - d_{max}) \cos n\Delta\Phi(s) \right\rangle_0 \right\rangle_{\Delta\Omega} \right]$$
$$= \sum_{n=0}^{2} \left\{ \left[ \left\langle E_0^{(n)}(R_0 + r(s))\theta(r(s) - d_{min})\theta(d_{max} - r(s)) \right\rangle_r + E_0^{(n)}(R_0 + d_{min}) \left\langle \theta(d_{min} - r(s)) \right\rangle_r \right. \right.$$
$$+ E_0^{(n)}(R_0 + d_{max}) \left\langle \theta(r(s) - d_{max}) \right\rangle_r \right] + \left[ \left\langle E_1^{(n)}(R_0 + r(s))\theta(r(s) - d_{min})\theta(d_{max} - r(s)) \right\rangle_r$$
$$\left. + E_1^{(n)}(R_0 + d_{min}) \left\langle \theta(d_{min} - r(s)) \right\rangle_r + E_0^{(n)}(R_0 + d_{max}) \left\langle \theta(r(s) - d_{max}) \right\rangle_r \right] \left\langle \sin\eta(s) \right\rangle_\eta \right\} \left\langle \left\langle \cos n\Delta\Phi(s) \right\rangle_{\Delta\Phi} \right\rangle_{\Delta\Omega}.$$
(B.27)

Then using Eq. (B.7) for the $r$-averages, we obtain

$$\left[ \left\langle E_0^{(0)}(R_0 + r(s))\theta(r(s) - d_{min})\theta(d_{max} - r(s)) \right\rangle_r + E_0^{(0)}(R_0 + d_{min}) \left\langle \theta(d_{min} - r(s)) \right\rangle_r \right.$$
$$\left. + E_0^{(0)}(R_0 + d_{max}) \left\langle \theta(r(s) - d_{max}) \right\rangle_r \right] = \frac{2l_B}{l_c^2} \frac{(1-\theta)^2}{(\kappa_D a K_1(\kappa_D a))^2}$$
$$\times \left[ \frac{1}{d\sqrt{2\pi}} \int_{d_{min}}^{d_{max}} dr K_0(\kappa_D(R_0 + r)) \exp\left(-\frac{r^2}{2d_r^2}\right) + \frac{K_0(\kappa_D(R_0 + d_{min}))}{2} \left(1 - \text{erf}\left(\frac{d_{min}}{d_r\sqrt{2}}\right)\right) \right. \quad \text{(B.28)}$$
$$\left. + \frac{K_0(\kappa_D(R_0 + d_{max}))}{2} \left(1 - \text{erf}\left(\frac{d_{max}}{d_r\sqrt{2}}\right)\right) \right].$$

$$\left[\left\langle E_0^{(n)}(R_0+r(s))\theta(r(s)-d_{\min})\theta(d_{\max}-r(s))\right\rangle_r + E_0^{(n)}(R_0+d_{\min})\left\langle\theta(d_{\min}-r(s))\right\rangle_r\right.$$

$$\left.+E_0^{(n)}(R_0+d_{\max})\left\langle\theta(r(s)-d_{\max})\right\rangle_r\right] = \frac{4l_B}{l_c^2}\frac{(-1)^n \zeta_n^2}{(\kappa_n a K_n'(\kappa_n a))^2}$$

$$\times\left[\frac{1}{d\sqrt{2\pi}}\int_{d_{\min}}^{d_{\max}} dr K_0(\kappa_n(R_0+r))\exp\left(-\frac{r^2}{2d_r^2}\right) + \frac{K_0(\kappa_n(R_0+d_{\min}))}{2}\left(1-\mathrm{erf}\left(\frac{d_{\min}}{d_r\sqrt{2}}\right)\right)\right.\quad \text{for } n\neq 0$$

$$\left.+\frac{K_0(\kappa_n(R_0+d_{\max}))}{2}\left(1-\mathrm{erf}\left(\frac{d_{\max}}{d_r\sqrt{2}}\right)\right)\right]$$

(B.29)

Similarly, we find

$$\left[\left\langle E_1^{(n)}(R_0+r(s))\theta(r(s)-d_{\min})\theta(d_{\max}-r(s))\right\rangle_r + E_1^{(n)}(R_0+d_{\min})\left\langle\theta(d_{\min}-r(s))\right\rangle_r\right.$$

$$\left.+E_1^{(n)}(R_0+d_{\max})\left\langle\theta(r(s)-d_{\max})\right\rangle_r\right] = \frac{4l_B n^2}{l_c^2}\frac{\overline{g}}{\kappa_n}\frac{(-1)^n \zeta_n^2}{(\kappa_n a K_n'(\kappa_n a))^2}$$

$$\times\left[\frac{1}{d\sqrt{2\pi}}\int_{d_{\min}}^{d_{\max}} dr K_1(\kappa_n(R_0+r))\exp\left(-\frac{r^2}{2d_r^2}\right) + \frac{K_1(\kappa_n(R_0+d_{\min}))}{2}\left(1-\mathrm{erf}\left(\frac{d_{\min}}{d_r\sqrt{2}}\right)\right)\right.\quad \text{(B.30)}$$

$$\left.+\frac{K_1(\kappa_n(R_0+d_{\max}))}{2}\left(1-\mathrm{erf}\left(\frac{d_{\max}}{d_r\sqrt{2}}\right)\right)\right]$$

The $\eta$-average yields

$$\left\langle \sin\eta(s)\right\rangle_\eta = \sin\eta_0 \exp\left(-\frac{d_\eta^2}{2}\right) \tag{B.31}$$

For the $\Delta\Phi$-average we can write

$$\left\langle\left\langle\cos(n\Delta\Phi(s))\right\rangle_{\Delta\Phi}\right\rangle_{\Delta\Omega} = \left\langle\cos(n\Delta\Phi_0(s))\right\rangle_{\Delta\Omega}\left\langle\cos(n\delta\Phi(s))\right\rangle_{\Delta\Phi} - \left\langle\sin(n\Delta\Phi_0(s))\right\rangle_{\Delta\Omega}\left\langle\sin(n\delta\Phi(s))\right\rangle_{\Delta\Phi}$$

$$= \left\langle\cos(n\Delta\Phi_0(s))\right\rangle_{\Delta\Omega}\left\langle\cos(n\delta\Phi(s))\right\rangle_{\Delta\Phi},$$

(B.32)

using the fact that $\left\langle\sin(n\Delta\Phi'(s))\right\rangle_{\Delta\Phi} = 0$. From Eq. (B.7) it is easy to show that

$$\left\langle\cos n\Delta\Phi'(s)\right\rangle_{\Delta\Phi} = \exp\left(-\frac{n^2 d_\Phi^2}{2}\right). \tag{B.33}$$

We can write for the ensemble average over $\Delta\Omega(s)$

$$\langle \cos n\Delta\Phi_0(s)\rangle_{\Delta\Omega}$$

$$= \frac{\int D\Delta\Omega(s)\cos\left(\Delta\bar{\Phi} + \frac{\sigma_H}{2\langle h\rangle}\int_{-L_B/2}^{L_B/2}\frac{\Delta\Omega(s')(s-s')}{|s-s'|}\exp\left(-\frac{|s-s'|}{\lambda_h}\right)\right)\exp\left(-\frac{\lambda_c^{(0)}}{4\langle h\rangle^2}\int_{-L_B/2}^{L_B/2}\Delta\Omega(s)^2 ds\right)}{\int D\Delta\Omega(s)\exp\left(-\frac{\lambda_c^{(0)}}{4\langle h\rangle^2}\int_{-L_B/2}^{L_B/2}\Delta\Omega(s)^2 ds\right)}.$$

(B.34)

Eq. (B.34) evaluates to

$$\langle \cos n\Delta\Phi_0(s)\rangle_{\Delta\Omega} = \cos\Delta\bar{\Phi}\exp\left(-\frac{\sigma_H \lambda_h}{4\lambda_c^{(0)}}\right). \tag{B.35}$$

**Expressions for $d_r$ $\theta_r$, $d_\eta$ and $d_\Phi$**

The starting expressions (which are derived from the Fourier transforms of Eq. (2.32) of the main text and the general expression for the averages Eq. (B.8)) for $d_r$ $\theta_r$, $d_\eta$ and $d_\Phi$ are

$$d_r^2 = \frac{1}{2\pi}\int_{-\infty}^{\infty}\frac{dk}{l_p^b k^4/2 + k_r}, \quad \theta_r^2 = \frac{1}{2\pi}\int_{-\infty}^{\infty}\frac{k^2 dk}{l_p^b k^4/2 + k_r}, \quad d_\eta^2 = \frac{1}{2\pi}\int_{-\infty}^{\infty}\frac{dk}{l_p^b k^2/2 + k_\eta} \quad \text{and}$$

$$d_\Phi^2 = \frac{1}{2\pi}\int_{-\infty}^{\infty}\frac{dk}{l_p^h k^2/2 + k_\eta}. \tag{B.36}$$

All of these integrals can be evaluated yielding

$$d_r^2 = \frac{1}{k_r}\left(\frac{k_r}{l_p^b}\right)^{1/4}\frac{1}{2^{5/4}}, \quad \theta_r^2 = \frac{1}{k_r}\left(\frac{k_r}{l_p^b}\right)^{3/4}\frac{1}{2^{5/4}}, \quad d_\eta^2 = \frac{1}{k_\eta}\left(\frac{k_\eta}{l_p^b}\right)^{1/2}\frac{1}{2^{1/2}} \quad \text{and} \quad d_\Phi^2 = \frac{1}{k_\Phi}\left(\frac{k_\Phi}{l_p^h}\right)^{1/2}\frac{1}{2^{1/2}}.$$

(B.37)

It is also useful to define the lengths

$$\lambda_\eta = \left(\frac{l_p^b}{2k_\eta}\right)^{1/2} \quad \text{and} \quad \lambda_h = \left(\frac{l_p^b}{2k_\Phi}\right)^{1/2}. \tag{B.38}$$

Here, it is automatically assumed that the parameter $\lambda_h$ can be used for both the thermal fluctuations and in the variational trial functional (Eq. (B.22)) for the intrinsic DNA distortions. A more rigorous approach where $d_\Phi^2$ and $\lambda_h$ are considered independent of each other and minimized separately [**Error! Bookmark not defined.**], justifies this assumption. In Eq.(B.9), we can evaluate the averages

$$\frac{1}{2}\int_{-L/2}^{L/2} ds\left((k_0-k_r)\langle r(s)^2\rangle_0 - k_\Phi\langle\delta\Phi(s)^2\rangle_0 - k_\eta\langle\delta\eta(s)^2\rangle_0\right)$$

$$= \frac{d_r^2 L}{(d_{max}-d_{min})^{8/3}(l_p^b)^{1/3}} - \frac{L}{d_r^{2/3}2^{8/3}(l_p^b)^{1/3}} - \frac{L}{4\lambda_\eta} - \frac{L}{4\lambda_h}. \tag{B.39}$$

**Calculation of** $\ln Z_{eff}$

We utilize the fact that

$$\frac{\partial \ln Z_{eff}}{\partial k_\eta} = -\frac{L}{2}\langle\delta\eta(s)^2\rangle = -\frac{d_\eta^2 L}{2}, \tag{B.40}$$

$$\frac{\partial \ln Z_{eff}}{\partial k_r} = -\frac{L}{2}\langle r(s)^2\rangle = -\frac{d_r^2 L}{2}, \tag{B.41}$$

$$\frac{\partial \ln Z_{eff}}{\partial k_\Phi} = -\frac{L}{2}\langle\delta\Phi(s)^2\rangle = -\frac{d_\Phi^2 L}{2}. \tag{B.42}$$

Using Eq. (B.37), all the expressions in Eqs. (B.40),(B.41) and (B.42) can be rewritten as

$$\frac{\partial \ln Z_{eff}}{\partial k_\eta} = -\frac{L}{2\sqrt{2}k_\eta^{1/2}}\left(\frac{1}{l_p^b}\right)^{1/2} \quad \frac{\partial \ln Z_{eff}}{\partial k_r} = -\frac{L}{k_r^{3/4}}\left(\frac{1}{l_p^b}\right)^{1/4}\frac{1}{2^{9/4}} \quad \frac{\partial \ln Z_{eff}}{\partial k_\Phi} = -\frac{L}{2\sqrt{2}k_\Phi^{1/2}}\left(\frac{1}{l_p^h}\right)^{1/2}$$

$$\tag{B.43}$$

We can then integrate up all of these expressions

$$-\ln Z_{eff} = \frac{L}{\sqrt{2}}\left(\frac{k_\eta}{l_p^b}\right)^{1/2} + \frac{L}{\sqrt{2}}\left(\frac{k_\Phi}{l_p^h}\right)^{1/2} + \left(\frac{k_r}{l_p^b}\right)^{1/4}\frac{4L}{2^{9/4}} + \Theta_0$$

$$= \frac{4L}{2^{8/3}d_r^{2/3}(l_p^b)^{1/3}} + \frac{L}{2\lambda_h} + \frac{L}{2\lambda_\eta} + \Theta_0, \tag{B.44}$$

where $\Theta_0$ is an ill-defined constant of integration that can be discarded. Then by combining Eqs (B.9),(B.10),(B.15),(B.16), (B.23),(B.27), (B.28),(B.29),(B.30), (B.31),(B.32),(B.35),(B.38),(B.39) and (B.44), setting $\lambda_h^* = \frac{\lambda_h}{2}\left(1+\frac{\lambda_c}{l_p^h}\right)$, $\lambda_c = \frac{\sigma_H \lambda_c^{(0)} l_p^h}{\lambda_c^{(0)}+l_p^h} + (1-\sigma_H)l_p^h$, and setting $d_{max}=\infty$, we arrive at Eqs. (2.33)-(2.37) of the main text, using the form factor for a DNA like charge distribution Eq. (A.43). By calculating $d_{max}$ from our numerical results, we find that $d_{max}$ is large enough to set it to infinity in all the expressions for the average electrostatic and bending energies to high accuracy.

## Appendix C: Minimization of Free Energy

Here we write down a set of equations for $d_r$, $\lambda_h$, $\lambda_\eta$, $R_0$, $\Delta\bar{\Phi}$ and $\eta_0$, the solution to which minimizes the Free energy function (Eq. (2.33) of the main text) i.e.

$$\frac{\partial F_T}{\partial d_r}=0, \quad \frac{\partial F_T}{\partial \lambda_h}=0, \quad \frac{\partial F_T}{\partial \lambda_\eta}=0, \quad \frac{\partial F_T}{\partial R_0}=0, \quad \frac{\partial F_T}{\partial \Delta\bar{\Phi}}=0 \text{ and } \frac{\partial F_T}{\partial \eta_0}=0. \tag{C.1}$$

The minimization conditions (Eq. (C.1)) with Eqs. (2.33)-(2.37) of the main text then yields the following set of equations (in the order given by Eq. (C.1))

$$\frac{d_r}{2^{5/3}\left(l_p^b\right)^{1/3}}\left[\frac{1}{d_r^{8/3}}-\frac{2^{8/3}}{(d_{max}-d_{min})^{8/3}}\right]$$

$$=\frac{l_p^b}{d_r R_0^2}\ell(R_0,d_r;a)\left[\frac{3}{2}-2\cos\eta_0\exp\left(-\frac{\lambda_\eta}{2l_p^b}\right)+\frac{\cos 2\eta_0}{2}\exp\left(-\frac{2\lambda_\eta}{l_p^b}\right)\right]$$

$$-\frac{1}{d_r}\frac{2l_B(1-\theta)^2}{l_c^2}\frac{m_0(\kappa_D R_0,\kappa_D d_r,(R_0-2a)/d_r)}{(\kappa_D a K_1(\kappa_D a))^2}-\frac{4l_B}{d_r l_c^2}\frac{[\zeta_1(f_1,f_2,\theta)]^2}{(\kappa_1 a K_1'(\kappa_1 a))^2}\cos\Delta\bar{\Phi}\exp\left(-\frac{\lambda_h^*}{2\lambda_c}\right)$$

$$\times\left[m_0(\kappa_1 R_0,\kappa_1 d_r,(R_0-2a)/d_r)+\bar{g}/\kappa_1\sin\eta_0\exp\left(-\frac{\lambda_\eta}{2l_p^b}\right)m_1(\kappa_1 R_0,\kappa_1 d_r,(R_0-2a)/d_r)\right]$$

$$+\frac{4l_B}{d_r l_c^2}\frac{[\zeta_2(f_1,f_2,\theta)]^2}{(\kappa_2 a K_2'(\kappa_2 a))^2}\cos 2\Delta\bar{\Phi}\exp\left(-\frac{2\lambda_h^*}{\lambda_c}\right)$$

$$\times\left[m_0(\kappa_2 R_0,\kappa_2 d_r,(R_0-2a)/d_r)+4\bar{g}/\kappa_2\sin\eta_0\exp\left(-\frac{\lambda_\eta}{2l_p^b}\right)m_1(\kappa_2 R_0,\kappa_2 d_r,(R_0-2a)/d_r)\right]$$

$$+\frac{2l_B}{d_r l_c^2}\sum_{n=-\infty}^{\infty}\frac{m_{img}(n,\kappa_n R_0,\kappa_n d_r,(R_0-2a)/d_r;a)}{(\kappa_n a K_n'(\kappa_n a))^2}[\zeta_n(f_1,f_2,\theta)]^2, \tag{C.2}$$

$$\lambda_h^*=\sqrt{\frac{(l_p^h+\lambda_c)^2}{8l_p^h\left(a_1(\eta_0,R_0,\lambda_\eta,d_r)\cos\Delta\bar{\Phi}\exp\left(-\frac{\lambda_h^*}{2\lambda_c}\right)-4a_2(\eta_0,R_0,\lambda_\eta,d_r)\cos 2\Delta\bar{\Phi}\exp\left(-\frac{2\lambda_h^*}{\lambda_c}\right)\right)}}, \tag{C.3}$$

$$\lambda_\eta = \frac{R_0}{2}\Bigg( f(R_0, d_r; a)\bigg( \cos\eta_0 \exp\bigg(-\frac{\lambda_\eta}{2l_p^b}\bigg) - \cos 2\eta_0 \exp\bigg(-\frac{2\lambda_\eta}{l_p^b}\bigg) \bigg)$$

$$+ \frac{\sin\eta_0 R_0^2}{l_p^b} \frac{2l_B \bar{g}}{l_c^2 \kappa_1} \frac{[\zeta_1(f_1, f_2, \theta)]^2}{(\kappa_1 a K_1'(\kappa_1 a))^2} \cos\Delta\bar{\Phi} \exp\bigg(-\frac{\lambda_h^*}{2\lambda_c}\bigg) \exp\bigg(-\frac{\lambda_\eta}{2l_p^b}\bigg) g_1(\kappa_1 R_0, \kappa_1 d_r, (R_0-2a)/d_r)$$

$$- \frac{\sin\eta_0 R_0^2}{l_p^b} \frac{8l_B \bar{g}}{l_c^2 \kappa_2} \frac{[\zeta_2(f_1, f_2, \theta)]^2}{(\kappa_2 a K_2'(\kappa_2 a))^2} \cos 2\Delta\bar{\Phi} \exp\bigg(-\frac{2\lambda_h^*}{\lambda_c}\bigg) \exp\bigg(-\frac{\lambda_\eta}{2l_p^b}\bigg) g_1(\kappa_2 R_0, \kappa_2 d_r, (R_0-2a)/d_r) \Bigg)^{-1/2},$$

(C.4)

$$0 = -\frac{8}{3} \frac{d_r^2}{(d_{max}-d_{min})^{11/3} (l_p^b)^{1/3}} \frac{d(d_{max}-d_{min})}{dR_0}$$

$$- \frac{l_p^b}{R_0^3} h(R_0, d_r; a) \bigg[ \frac{3}{2} - 2\cos\eta_0 \exp\bigg(-\frac{\lambda_\eta}{2l_p^b}\bigg) + \frac{\cos 2\eta_0}{2} \exp\bigg(-\frac{2\lambda_\eta}{l_p^b}\bigg) \bigg]$$

$$+ \frac{2l_B(1-\theta)^2}{R_0 l_c^2} \frac{q_0(\kappa_D R_0, \kappa_D d_r, (R_0-2a)/d_r)}{(\kappa_D a K_1(\kappa_D a))^2}$$

$$- \frac{4l_B}{R_0 l_c^2} \frac{[\zeta_1(f_1, f_2, \theta)]^2}{(\kappa_1 a K_1'(\kappa_1 a))^2} \cos\Delta\bar{\Phi} \exp\bigg(-\frac{\lambda_h^*}{2\lambda_c}\bigg)$$

$$\times \bigg[ q_0(\kappa_1 R_0, \kappa_1 d_r, (R_0-2a)/d_r) + \bar{g}/\kappa_1 \sin\eta_0 \exp\bigg(-\frac{\lambda_\eta}{2l_p^b}\bigg) q_1(\kappa_1 R_0, \kappa_1 d_r, (R_0-2a)/d_r) \bigg]$$

$$+ \frac{4l_B}{R_0 l_c^2} \frac{[\zeta_2(f_1, f_2, \theta)]^2}{(\kappa_2 a K_2'(\kappa_2 a))^2} \cos 2\Delta\bar{\Phi} \exp\bigg(-\frac{2\lambda_h^*}{\lambda_c}\bigg)$$

$$\times \bigg[ q_0(\kappa_2 R_0, \kappa_2 d_r, (R_0-2a)/d_r) + 4\bar{g}/\kappa_2 \sin\eta_0 \exp\bigg(-\frac{\lambda_\eta}{2l_p^b}\bigg) q_1(\kappa_2 R_0, \kappa_2 d_r, (R_0-2a)/d_r) \bigg]$$

$$+ \frac{2l_B}{R_0 l_c^2} \sum_{n=-\infty}^{\infty} \frac{q_{img}(n, \kappa_n R_0, \kappa_n d_r, (R_0-2a)/d_r; a)}{(\kappa_n a K_n'(\kappa_n a))^2} [\zeta_n(f_1, f_2, \theta)]^2,$$

(C.5)

$$\cos\Delta\bar{\Phi} = \frac{a_1(\eta_0, R_0, \lambda_\eta, d_r)}{4a_2(\eta_0, R_0, \lambda_\eta, d_r)} \exp\bigg(\frac{3\lambda_h^*}{2\lambda_c}\bigg),$$

(C.6)

$$0 = -\frac{8}{3}\frac{d_r^2}{(d_{max}-d_{min})^{11/3}(l_p^b)^{1/3}}\frac{d(d_{max})}{d\eta_0}$$

$$+\frac{l_p^b}{R_0^2}f(R_0,d_r;a)\left[2\sin\eta_0\exp\left(-\frac{\lambda_\eta}{2l_p^b}\right)-\sin 2\eta_0\exp\left(-\frac{2\lambda_\eta}{l_p^b}\right)\right]$$

$$-\cos\eta_0\frac{4l_B\bar{g}}{l_c^2\kappa_1}\frac{[\zeta_1(f_1,f_2,\theta)]^2}{(\kappa_1 aK_1'(\kappa_1 a))^2}\cos\Delta\bar{\Phi}\exp\left(-\frac{\lambda_h^*}{2\lambda_c}\right)\exp\left(-\frac{\lambda_\eta}{2l_p^b}\right)g_1(\kappa_1 R_0,\kappa_1 d_r,(R_0-2a)/d_r)$$

$$+\cos\eta_0\frac{16l_B\bar{g}}{l_c^2\kappa_2}\frac{[\zeta_2(f_1,f_2,\theta)]^2}{(\kappa_2 aK_2'(\kappa_2 a))^2}\cos 2\Delta\bar{\Phi}\exp\left(-\frac{2\lambda_h^*}{\lambda_c}\right)\exp\left(-\frac{\lambda_\eta}{2l_p^b}\right)g_1(\kappa_2 R_0,\kappa_2 d_r,(R_0-2a)/d_r)$$

(C.7)

The functions given in Eq. (C.2)-(C.7) are defined as

$$a_n(\eta_0,R_0,\lambda_\eta,d_r) = \frac{4l_B}{l_c^2}\frac{[\zeta_n(f_1,f_2,\theta)]^2}{(\kappa_n aK_n'(\kappa_n a))^2}$$

$$\times\left[g_0(\kappa_n R_0,\kappa_n d_r,(R_0-2a)/d_r) + n^2 g/\kappa_n \sin\eta_0\exp\left(-\frac{\lambda_\eta}{2l_p^b}\right)g_1(\kappa_n R_0,\kappa_n d_r,(R_0-2a)/d_r)\right]$$

(C.8)

$$m_j(\kappa_n R_0,\kappa_n d_r,(R_0-2a)/d_r) = d_r\frac{\partial g_j(\kappa_n R_0,\kappa_n d_r,(R_0-2a)/d_r)}{\partial d_r}$$

$$= \frac{1}{\sqrt{2\pi}}\int_{(2a-R_0)/d_r}^{\infty}dy(y^2-1)\exp\left(-\frac{y^2}{2}\right)K_j(\kappa_n(R_0+yd_r))$$

$$+\frac{(R_0-2a)}{d_r\sqrt{2\pi}}K_j(2k_n a)\exp\left(-\frac{(R_0-2a)^2}{2d_r^2}\right),$$

(C.9)

$$m_{img}(n,\kappa_n R_0,\kappa_n d_r,(R_0-2a)/d_r;a) = d_r\frac{\partial g_{img}(\kappa_j R_0,\kappa_j d_r,(R_0-2a)/d_r)}{\partial d_r}$$

$$= -\frac{1}{\sqrt{2\pi}}\sum_{j=-\infty}^{\infty}\int_{\frac{(2a-R_0)}{d_r}}^{\infty}dy K_{n-j}(\kappa_n R_0+y\kappa_n d_r)K_{n-j}(\kappa_n R_0+y\kappa_n d_r)\frac{I_j'(\kappa_n a)}{K_j'(\kappa_n a)}(y^2-1)\exp\left(-\frac{y^2}{2}\right)$$

$$-\frac{1}{\sqrt{2\pi}}\frac{(R_0-2a)}{d_r}\sum_{j=-\infty}^{\infty}K_{n-j}(2\kappa_n a)K_{n-j}(2\kappa_n a)\frac{I_j'(\kappa_n a)}{K_j'(\kappa_n a)}\exp\left(-\frac{1}{2}\frac{(R_0-2a)^2}{d_r^2}\right),$$

(C.10)

$$\ell(R_0, d_r; a) = d_r \frac{\partial f(R_0, d_r; a)}{\partial d_r}$$

$$= \frac{R_0^2}{\sqrt{2\pi}} \int_{(2a-R_0)/d_r}^{\infty} dy \frac{(y^2-1)}{(R_0+d_r y)^2} \exp\left(-\frac{y^2}{2}\right) + \frac{1}{\sqrt{2\pi}} \frac{(R_0-2a)}{d_r} \left(\frac{R_0}{2a}\right)^2 \exp\left(-\frac{1}{2}\left(\frac{R_0-2a}{d_r}\right)^2\right),$$

(C.11)

$$q_j(\kappa_n R_0, \kappa_n d_r, (R_0-2a)/d_r) = R_0 \frac{\partial g_j(\kappa_n R_0, \kappa_n d_r, (R_0-2a)/d_r)}{\partial R_0}$$

$$= \frac{1}{\sqrt{2\pi}} \frac{R_0}{d_r} \int_{-1}^{\infty} dx K_j\left(\kappa_n(R_0+x(R_0-2a))\right) \exp\left(-\frac{x^2}{2}\left(\frac{R_0-2a}{d_r}\right)^2\right)$$

$$+ \frac{1}{\sqrt{2\pi}} \frac{(R_0-2a)}{d_r} \kappa_n R_0 \int_{-1}^{\infty} dx K'_j\left(\kappa_n(R_0+x(R_0-2a))\right)(1+x) \exp\left(-\frac{x^2}{2}\left(\frac{R_0-2a}{d_r}\right)^2\right)$$

(C.12)

$$- \frac{1}{\sqrt{2\pi}} \frac{(R_0-2a)^2 R_0}{d_r^3} \int_{-1}^{\infty} dx x^2 K_j\left(\kappa_n(R_0+x(R_0-2a))\right) \exp\left(-\frac{x^2}{2}\left(\frac{R_0-2a}{d_r}\right)^2\right)$$

$$- \frac{R_0}{d_r \sqrt{2\pi}} K_j(2k_n a) \exp\left(-\frac{1}{2}\left(\frac{R_0-2a}{d_r}\right)^2\right),$$

$$q_{img}(n, \kappa_n R_0, \kappa_n d_r, (R_0-2a)/d_r; a) = R_0 \frac{\partial g_{img}(n, \kappa_n R_0, \kappa_n d_r, (R_0-2a)/d_r; a)}{\partial R_0}$$

$$= -\frac{1}{\sqrt{2\pi}} \frac{R_0}{d_r} \sum_{j=-\infty}^{\infty} \int_{-1}^{\infty} dx K_{n-j}(\kappa_n R_0 + x\kappa_n(R_0-2a)) K_{n-j}(\kappa_n R_0 + x\kappa_n(R_0-2a)) \frac{I'_j(\kappa_n a)}{K'_j(\kappa_n a)}$$

$$\times \exp\left(-\frac{x^2}{2}\left(\frac{R_0-2a}{d_r}\right)^2\right) - \frac{2}{\sqrt{2\pi}} \frac{(R_0-2a)\kappa_n R_0}{d_r} \sum_{j=-\infty}^{\infty} \frac{I'_j(\kappa_n a)}{K'_j(\kappa_n a)} \int_{-1}^{\infty} dx K'_{n-j}\left(\kappa_n(R_0+x(R_0-2a))\right)$$

$$\times K_{n-j}\left(\kappa_n(R_0+x(R_0-2a))\right)(1+x) \exp\left(-\frac{x^2}{2}\left(\frac{R_0-2a}{d_r}\right)^2\right)$$

$$+ \frac{1}{\sqrt{2\pi}} \frac{(R_0-2a)^2 R_0}{d_r^3} \sum_{j=-\infty}^{\infty} \frac{I'_j(\kappa_n a)}{K'_j(\kappa_n a)} \int_{-1}^{\infty} dx x^2 K_{n-j}\left(\kappa_n(R_0+x(R_0-2a))\right)$$

$$\times K_{n-j}\left(\kappa_n(R_0+x(R_0-2a))\right) \exp\left(-\frac{x^2}{2}\left(\frac{R_0-2a}{d_r}\right)^2\right)$$

$$+ \frac{1}{\sqrt{2\pi}} \frac{R_0}{d_r} \sum_{j=-\infty}^{\infty} K_{n-j}(2\kappa_n a) K_{n-j}(2\kappa_n a) \frac{I'_j(\kappa_n a)}{K'_j(\kappa_n a)} \exp\left(-\frac{1}{2} \frac{(R_0-2a)^2}{d_r^2}\right),$$

(C.13)

$$h(R_0, d_r; a) = f(R_0, d_r; a) - \frac{2a}{d_r} \frac{1}{\sqrt{2\pi}} \int_{-1}^{\infty} dx \frac{\left(1 - (1 - 2a/R_0)x\right)}{\left(1 + (1 - 2a/R_0)x\right)^3} \exp\left(-\frac{x^2}{2}\left(\frac{R_0 - 2a}{d_r}\right)^2\right)$$

$$+ \frac{R_0(R_0 - 2a)^2}{d_r^3} \frac{1}{\sqrt{2\pi}} \int_{-1}^{\infty} dx \frac{x^2}{\left(1 + (1 - 2a/R_0)x\right)^2} \exp\left(-\frac{x^2}{2}\left(\frac{R_0 - 2a}{d_r}\right)^2\right) \quad \text{(C.14)}$$

$$+ \frac{1}{\sqrt{2\pi}} \left(\frac{R_0}{2a}\right)^2 \left(\frac{R_0}{d_r}\right) \exp\left(-\frac{(R_0 - 2a)^2}{2d_r^2}\right) - \frac{1}{2}\left(\frac{R_0}{2a}\right)^2 \left(1 - \text{erf}\left(\frac{R_0 - 2a}{d_r\sqrt{2}}\right)\right).$$